\documentclass[12pt,onecolumn]{IEEEtran}
\usepackage[utf8]{inputenc}
\usepackage{graphicx}
\usepackage{amsmath,amssymb,amsfonts}
\usepackage{subfigure}
\usepackage{caption}
\usepackage{xcolor}
\usepackage{etoolbox}
\usepackage{mathtools}
\usepackage{dsfont}
\usepackage{textcomp}
\usepackage{cite}
\usepackage{bm}
\usepackage{titlesec}
\usepackage[margin=1in]{geometry}
\usepackage[ruled]{algorithm2e}
\usepackage[section]{placeins}
\usepackage{setspace}

\title{Toward 6G Sidelink Reliability: MAC PRR Modeling for NR Mode 2 SPS and ns-3 Validation}
\author{\text{Liu Cao},~\IEEEmembership{Member,~IEEE}, Zhaoyu Liu, \text{Lyutianyang Zhang},~\IEEEmembership{Member,~IEEE}

\thanks{

Liu Cao, Zhaoyu Liu are with both City University of Hong Kong (Dongguan), Dongguan, China, and City University of Hong Kong, Hong Kong (email: \{liu.cao,72515198\}@cityu-dg.edu.cn). Lyutianyang Zhang is with the School of Microelectronics and Communication Engineering, Chongqing University, Chongqing, China (email: zhanglyutianyang@cqu.edu.cn). (\emph{Corresponding author: Lyutianyang Zhang, Liu Cao})
}

\thanks{This work is partly supported by the Guangdong-Hong Kong Universities ``1+1+1" Joint Research Collaboration Scheme. This work is also partly supported by the Youth Innovation Talent Project of Guangdong Provincial Universities (Grant No. 2025KQNCX17).}

}

\begin{document}

\maketitle


\begin{abstract}
5G New Radio (NR) Sidelink (SL) Mode 2 has enabled decentralized, infrastructure-less direct communications which is evolving to serve reliability-critical services in 6G SL.
Particularly, the channel access in NR SL Mode 2 relies on the Sensing-based Semi-Persistent Scheduling (SPS) whose key features significantly influence the packet reception ratio (PRR). While SPS has been widely studied, existing analytical models typically abstract or omit several NR-specific SPS features that are standardized in the 3rd Generation Partnership Project (3GPP), limiting their ability to explain how SPS parameters shape MAC collision dynamics and PRR. This paper develops an analytical MAC-layer PRR model for broadcast NR SL mode 2 by explicitly modeling SPS-driven MAC collision events. The model captures (i) Collisions caused by simultaneous resource reselection and (ii) Persistent collisions induced by resource keeping across resource reservation intervals (RRIs). Based on the event-level characterization, we derive closed-form expressions for the steady-state MAC collision probability and PRR. We further extend the analysis to incorporate under-explored SPS features, including the duplicate transmissions per RRI and the minimum resource-availability requirement for reselection, and quantify their impact on PRR in under-saturated regimes. The analytical results are validated using ns-3 simulations based on the 5G-LENA framework, showing close agreement under under-saturation and revealing deviations as the system approaches saturation. The proposed model provides mechanistic insight and design guidance of tuning the SPS parameters to improve 6G SL reliability.

\end{abstract}
\begin{IEEEkeywords}
6G Sidelink, NR Mode 2, SPS, MAC, ns-3, Validation.
\end{IEEEkeywords}
\vspace{2cm}

\section{Introduction}
\label{introduction}

 \IEEEPARstart{S}{idelink} (SL) is a 3rd Generation Partnership Project (3GPP) standardized technology that enables direct user-to-user communications, without the assistance of traditional cellular networks \cite{weerackody2023needs}. Current SL (Release 16) has evolved from the Proximity Services (ProSe) protocol, defined in Release 12 to support public safety applications, \cite{firstnet}. As of Release 16, SL operates in modes 1 and 2, supporting direct communication among multiple User Equipment (UEs). The key difference: in mode 1, the resources used by the UE-to-UE link are allocated by the base station (gNB), whereas in mode 2, resources are selected autonomously without any gNB coordination.

A promising application for SL is support for mission-critical data connectivity needs with Ultra-Reliable Low-Latency Communications (URLLC) requirements. These characteristics are helpful in various scenarios, such as Vehicle-to-Everything (V2X), various first-responder use, and industrial IoT, leading to continued interest in expanding SL scope. SL mode 2 supports the formation of local-area, ad-hoc networks without cellular infrastructure (e.g., remote areas or complex propagation environments such as underground, undersea, or airborne). All such applications also bring new technical challenges for network formation over SL, e.g., the necessity to re-architect key primitives such as neighbor discovery, authentication, and efficient channel acquisition \cite{weerackody2023needs}. Because of this, enhancements to the current SL continue to be an active topic in 3GPP Rel-18/Rel-19 with several study items \cite{3gpp.37.885}. 

This work focuses on performance evaluation of the NR SL mode 2 Semi Persistent Scheduling (SPS) protocol, a distributed media access control (MAC) scheme that was initially introduced in Long Term Evolution (LTE) SL mode 4 (Release 14) and subsequently adopted for 5G NR. SPS includes a {\em distributed sensing} feature that enables all UEs in a decentralized network to {\em monitor the share channel and determine which resources are in use, and avoid utilizing these (in-use resources) for upcoming message transmissions}. SL design has evolved from the original Rel-14 specifications to include 5G NR numerology, operate over multiple frequency bands such as FR2 (mmWave), and support more communication types (unicast and groupcast in addition to broadcast); correspondingly, this has led to updates in SPS design in terms of new parametrization to meet the reliability requirements faced with network reconfiguration and scalability. 

In this work, we focus on analyzing broadcast NR SL mode 2 that uses SPS for basic safety/status messaging in C-V2X scenarios. A significant set of prior work analyzing SPS performance has been primarily based on simulations \cite{nabil2018performance, cecchini2017ltev2vsim, mccarthy2019opencv2x, ali20213gpp, campolo20195g, bazzi2018study, chen2021performance, liu2023towards,xu2025enhanced,11054065}. For instance, \cite{cecchini2017ltev2vsim, mccarthy2019opencv2x, ali20213gpp} proposed different simulation frameworks to study LTE/NR V2X. While a custom framework was developed in Matlab in\cite{cecchini2017ltev2vsim}, \cite{mccarthy2019opencv2x} and \cite{ali20213gpp} built modules in Objective Modular Network Testbed in C++ (OMNeT++) and network Simulator 3 (ns-3), respectively. Both \cite{campolo20195g} and \cite{bazzi2018study} used the simulation framework from \cite{cecchini2017ltev2vsim} to explore the effects of varying the numerology and the physical (PHY)/MAC parameters on the Packet Reception Ratio (PRR) in V2X, respectively. In summary, none of these studies undertake any analytical modeling of SPS performance. A few efforts have been devoted to SL MAC and PHY analysis in recent years \cite{rehman2022analytical,rehman2023impact, 10533255,gonzalez2018analytical, gu2021performance, gu2022performance,11023842,10757976}, exploring the impact of system parameters and network conditions on SL performance. For example, \cite{rehman2022analytical,rehman2023impact} focused on the cross-layer (MAC + PHY) modeling of the C-V2X SPS in a vehicle platooning scenario and \cite{10533255} developed a standard-compliant NR V2X sidelink link-level simulator (MATLAB-based) that implements key NR V2X SL PHY features (e.g., two-stage SCI processing, PSCCH–PSSCH multiplexing, dedicated DMRS patterns, and standardized V2X channel models) following Release-16 specifications. An analytical VRU Awareness Probability (VAP) model was presented in \cite{11023842} to relate the packet delivery ratio (PDR) of NR sidelink mode 2 to urban VRU safety under realistic mobility and obstruction conditions; however, it abstracts sensing-based SPS operations and thus does not clarify how key mode 2 MAC/SPS parameters jointly determine collisions and PRR. A novel discrete-time Markov chain (DTMC) approach was presented in \cite{nba2020discrete} to calculate the average delay, collision probability and channel utilization in LTE SL mode 4. However, the proposed Markov-based SPS analysis incorporated only limited SPS parameters - the reselection counter and the selection window size. The authors in \cite{gonzalez2018analytical, gu2021performance} modeled the PRR and the access collision probability for LTE SL mode 4, respectively. However, they only focused on the legacy SPS protocol for LTE SL and failed to account for the new NR SL SPS features, such as the multiple packet transmission that significantly impacts the PRR. Meanwhile, most existing analytical models for the legacy SPS protocol may not provide insights into how SPS parameters relate to each other for determining the SL performance, such as PRR. However, understanding their mutual impacts is important for the reliability enhancement of future beyond 5G (B5G) and 6G SL services.

\begin{table}[t]
 \centering
 \caption{\small{Glossary
} 
}\label{tab: symbolGlossary
}
\resizebox{.55\textwidth}{!}{\begin{tabular}{ |c|c|c|c|c|c|c| } 
\hline
\textbf{Notation} & \textbf{Definition} \\
\hline
$T_{RRI}$ & Duration of Resource Reservation Interval (RRI), millisecond  \\
\hline
$R_{c,init}(T)$ & Initial reselection counter at time $T$\\
\hline
$R_c(T)$ & Reselection counter at time $T$\\
\hline
$p_k$ & Resource keeping probability \\
\hline
$T_{g}$ & The time the reference UE generates a packet\\
\hline
$N_{sc}$ & Number of subchannels\\
\hline
$t_s$ & Slot duration, millisecond  \\
\hline
$N_r$ & Total number of PRBs in each RRI \\
\hline
$R_c^{UE0}$ & UE 0's reselection counter\\
\hline
$R_{c,init}^{UE0}$ & UE 0's initial reselection counter\\
\hline
$P_{COL}$ & Total MAC collision probability \\\hline
$P_{COL,1}$ & The probability for Collision Event 1\\
\hline
$P_{COL,2}$ & The probability for Collision Event 2\\
\hline
$\pi_i$ & Probability that $R_c = i$ \\
\hline
$N_{UE}$ & Total number of UEs\\
\hline
$P_s(n)$ & The probability that $n$ UEs reselect within UE 0's selection window\\
\hline
$N_a$ & Number of available (unoccupied) PRBs in the selection window \\
\hline
$N_o$ & Number of occupied PRBs in the selection window \\
\hline
$N_c$ & Number of packets in an occupied PRB where collisions happen \\
\hline
$P_r(n)$ & The probability of $n$-fold collision given that $n$ UEs reselect\\
\hline
$P_{COL,2}^{E1}$ & $P_{COL,2}$ due to $E1$\\
\hline
$E1$ & $\text{Collision during } [T_1, T_1 + T_{RRI}], \text{and } R_{c,init}^{UE 0} = R_{c}^{UE 1} \text{ at } T = T_1 + T_{RRI}$\\
\hline
$P_{COL,2}^{E2}$ & $P_{COL,2}$ due to $E2$\\
\hline
$E2$ & $\text{Collision during } [T_1, T_1 + T_{RRI}], \text{and } R_{c,init}^{UE 0} < R_{c}^{UE 1} \text{ at } T = T_1 + T_{RRI}$\\
\hline
$P_{COL,2}^{E3}$ & $P_{COL,2}$ due to $E3$\\
\hline
$E3$ & $\text{Collision during } [T_1, T_1 + T_{RRI}], \text{and } R_{c,init}^{UE 0} > R_{c}^{UE 1} \text{ at } T = T_1 + T_{RRI}$\\
\hline
$P_{HD}$ & Probability that Half-duplex effects happen \\\hline
$PRR$ & Packet reception ratio \\\hline
$N_{Se}$ & Number of packets that each UE sends per RRI \\\hline
$X$ & Minimum proportion requirement of available PRBs\\
\hline
$d_{t,r}$ & Transmission range
 \\
\hline
\end{tabular}}
\vspace{-0.5cm}
\end{table} 


Motivated by the aforementioned limitations of existing analytical and simulation-based studies, we explicitly model the SPS-induced MAC collision processes that arise from asynchronous reselection and resource keeping, and we leverage this collision-event structure to derive an analytical model for the MAC-layer packet reception ratio (PRR). To the best of the authors' knowledge, this work provides the \emph{first complete analytical model of NR sidelink mode~2 MAC collision events} that is directly usable to derive MAC PRR. This is important because the MAC PRR under mode~2 SPS is fundamentally determined by the temporal evolution of MAC collision events across reservation intervals: collisions are not isolated one-shot events but can persist due to semi-persistent reservations and resource keeping. Without an explicit event-level model, it is hard to analytically quantify the steady-state MAC collision probability, and hence PRR, as a function of SPS parameters. The key contributions of this paper are
summarized as follows.
\begin{itemize}
    \item \textbf{MAC collision-event decomposition for NR SL mode 2.} We formally define and analyze two dominant SPS-driven MAC collision mechanisms in the UE selection window. This event structure is designed to reflect NR SPS operation and to expose which SPS behaviors create multi-RRI collision sequences.
    \item \textbf{Closed-form MAC collision probability and PRR model with key NR SPS parameters.} We derive an analytical model for the steady-state collision probability and PRR that connects reliability directly to SPS parameters, including the resource-keeping probability and asynchronous reselection timing, and produces numerically solvable closed-form expressions for MAC collision probability and PRR.
    \item \textbf{Model extensions for NR reliability features: duplication and minimum resource-availability constraint.} We extend the baseline model to quantify how the under-explored new SPS features affect the MAC PRR in under-saturated regimes.
    \item \textbf{Independent ns-3 validation and identification of the under-saturation operating region.} We implement and validate the analytical predictions using ns-3 (5G-LENA-based) simulations, demonstrating close agreement for under-saturated conditions and explaining deviations as the system approaches saturation—thereby clarifying when a pure-MAC analytical model is expected to be accurate and when PHY/sensing effects become significant. 
\end{itemize}

The rest of this paper is organized as follows. Sec. \ref{systemSetup} describes the system setup and defines the MAC collision events. Sec. \ref{SPSModel} provides the analytical model for the defined collision events and PRR for some MAC parameters, while Sec. \ref{extension} updates the model of collision probabilities and PRR to incorporate the number of duplicate transmissions. Sec. \ref{validation} then describes and shows results from the simulation regime used to validate the model.

\section{System setup}
\label{systemSetup}

Consider a typical SL mode 2 scenario where static UEs are randomly distributed in a local region such that the resulting network is fully connected. SL packet traffic sent from each UE is periodic and uses the sensing-based Semi-Persistent Scheduling (SPS) defined in \cite{3gpp.37.885,garcia2021tutorial} for channel access. An SL {\em resource pool}\footnote{A set of available time/frequency resources dedicated for SL TB transmissions.} comprised of a set of contiguous sub-channels\footnote{A subchannel is composed of a network maintainer configured number of resource blocks, which are in turn composed of 12 subcarriers whose bandwidth is defined by the subcarrier spacing.} in the frequency domain and slots in the time domain, enables transmission of (data) Transport Block (TB) which, for our purposes, constitutes a single SL Layer-2 packet\cite{3gpp.37.985}. This combination of one sub-channel and one slot to transmit a TB is a Physical Resource Block (PRB) that constitutes the unit resource in sensing-based SPS. We assume a perfect PHY-layer as befits a pure MAC analysis, i.e., any Layer-2 packet sent from a source is received perfectly by all intended receivers if not interfered with due to simultaneous transmission(s) by other UE sources.

\vspace{-0.5cm}
\subsection{Sensing-based SPS: Recap}

\begin{figure}[ht]
    \centering \includegraphics[width=.5\textwidth]{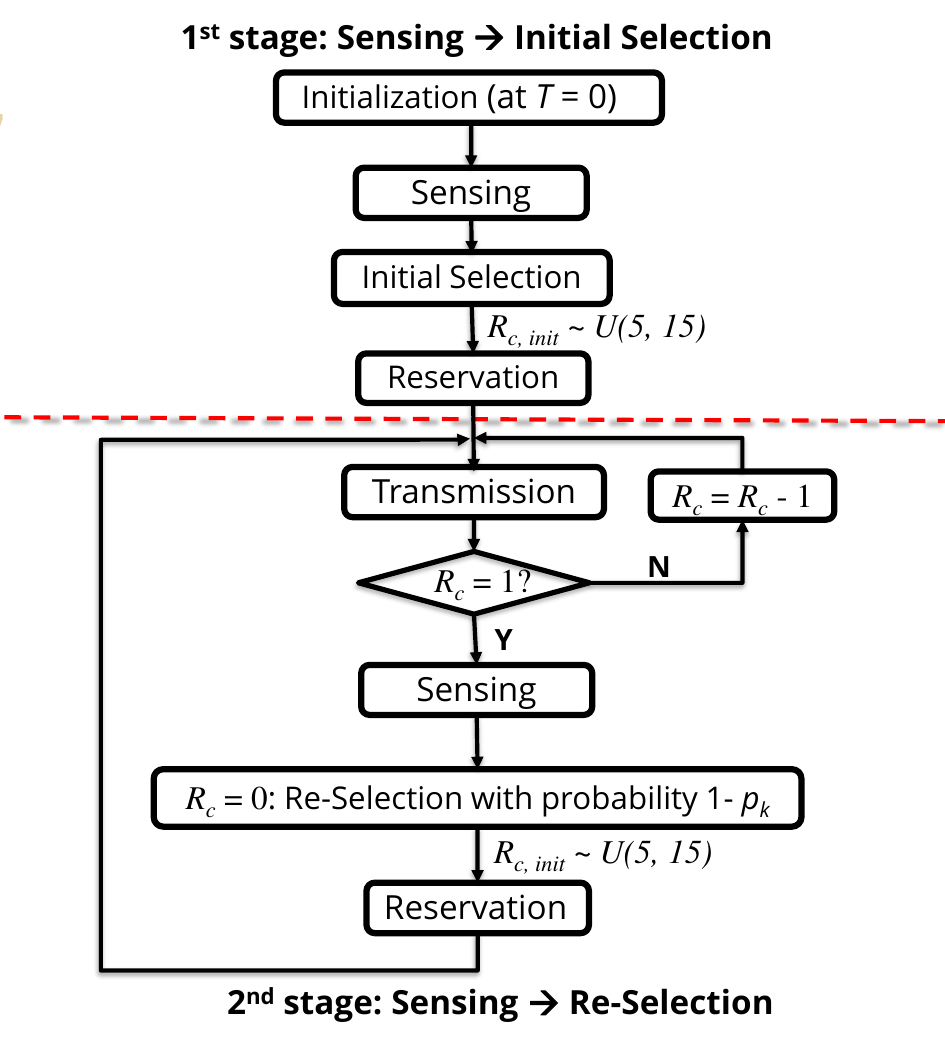}
    \caption{SPS MAC Procedure.}
    \vspace{-0.4cm}
    \label{fig:flowSPS}
\end{figure}

Sensing-based SPS \cite{3gpp.36.213, 3gpp.36.321} is a distributed MAC protocol for NR SL that seeks to achieve collision avoidance based on predictive resource reservation by UEs over a resource reservation interval (RRI), $T_{RRI}$. A flowchart of the SPS procedure is shown in Fig. \ref{fig:flowSPS}, which can be decomposed into two stages for purposes of understanding key aspects of protocol operation.

\begin{figure}[ht]
    \centering \includegraphics[width=.55\textwidth]{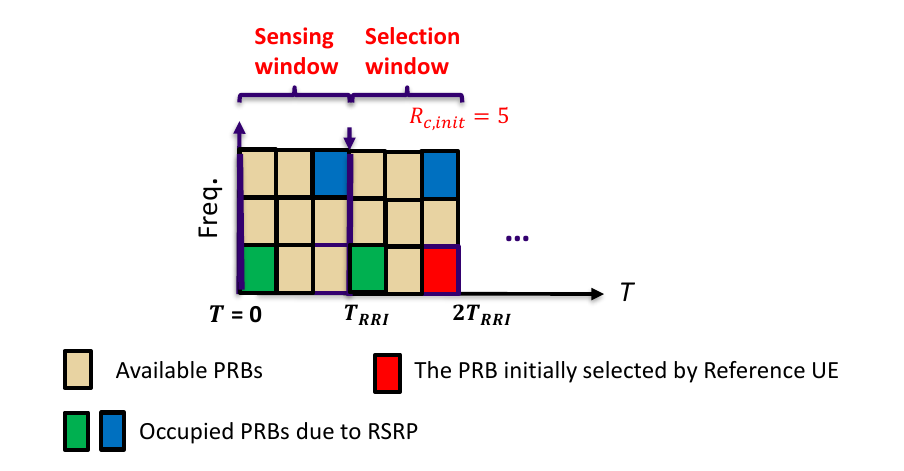}
    \caption{SPS Stage 1: Reference UE Case.}
    \label{fig:SPSStage1}
\end{figure}

{\em Stage 1: Sensing $\rightarrow$ Initial Selection.} This is the initial stage where a reference UE performs sensing-based initial selection for subsequent resource reservation.  A reference UE first determines the PRBs occupied by other UEs during the sensing window\footnote{The sensing window covers one or multiple $T_{RRI}$ defined in the 3GPP standard, depending on the upper layer setup.} via comparison to a reference signal received power (RSRP) threshold and then excludes those occupied PRBs for the subsequent reservation during selection window. The reference UE then \textbf{randomly} selects one among the remaining available PRBs in the selection window (for transmission of Layer-2 packet) and simultaneously chooses a {\em Reselection counter} $ R_{c, init}$, implying reservation of the initially selected PRB for subsequent $R_{c, init}$ transmissions. $R_{c, init}$ is selected as independent and identically distributed (i.i.d.) random variable from the uniform distribution ${U(5, 15)}$ \footnote{$R_{c, init}$ range varies if $T_{RRI}<100$ ms but always satisfies $R_{c, init} \in U(5, 15)$ for $T_{RRI} \geq 100$ ms per the 3GPP standard. Since $T_{RRI}$ is defined at the application layer, it is not the variable of interest for the MAC-layer analysis. For simplicity, we consider the SL applications with $T_{RRI} \geq 100$ ms, thus $R_{c, init} \in U(5, 15)$ range always holds.}. The counter is decremented by 1 after each $T_{RRI}$. An example is illustrated in Fig. \ref{fig:SPSStage1}.

\begin{figure}[ht]
    \centering \includegraphics[width=.6\textwidth]{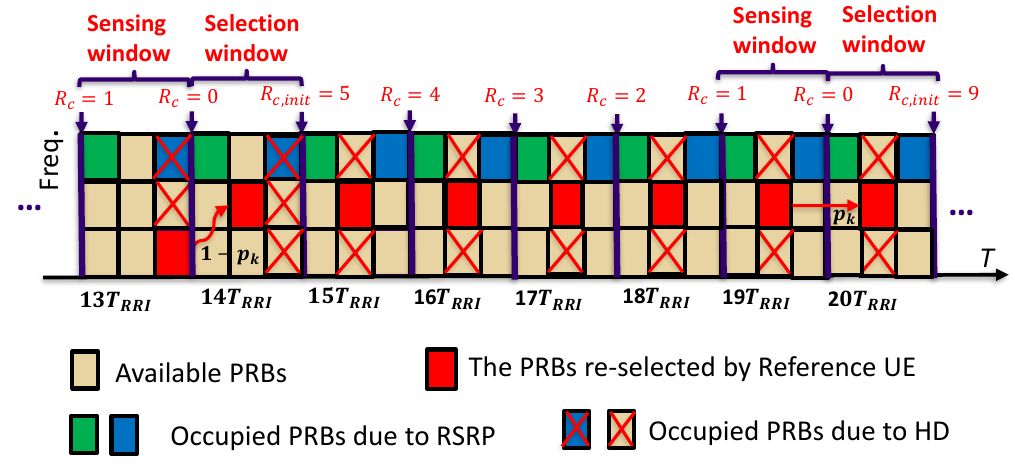}
    \caption{SPS Stage 2: Reference UE Case.}
    \label{fig:SPSStage2}
\end{figure}

\
{\em Stage 2: Sensing $\rightarrow$ Reselection.}
 In this stage, the reference UE starts packet transmissions, counts down the reselection counter until $R_c = 0$ and then enters reselection of PRBs:
 \begin{itemize}
    \item {\em Exclusion of occupied PRBs during the sensing window.}\\ 
    Since SL UEs are typically half-duplex (i.e., incapable of transmitting and receiving simultaneously), the reference UE cannot detect the RSRP of PRBs in the same slot when it transmits. Thus,  in the sensing window for reselection, reference UE will exclude all `occupied' PRBs due to not only exceeding the RSRP threshold but also the half-duplex constraint. 
    \item {\em Reselection.}\\
    When the reference UE's $R_c = 0$, it randomly \textbf{re-selects} one of the available PRBs with probability $1-p_k$ during the selection window while persisting with the prior reserved PRB selection with probability $p_k$.  $p_k \; \in \; [0, 0.8]$ is the {\em resource keeping} parameter specified in 3GPP standard and is identical for all UEs in the network. Such a reselection cycle is repeated in Stage 2 and is illustrated via example in Fig. \ref{fig:SPSStage2}. Since UE performs the process of sensing and initial selection only once during the $1^{st}$ stage while repeating the process of sensing and reselection during the $2^{nd}$ stage, the packet reception ratio (PRR) in the long term (i.e., the steady state) is determined by the $2^{nd}$ stage. We thereby develop the PRR model based on the events in the $2^{nd}$ stage.  \end{itemize}

\begin{figure}[ht]
    \centering
    \includegraphics[width=.55\textwidth]{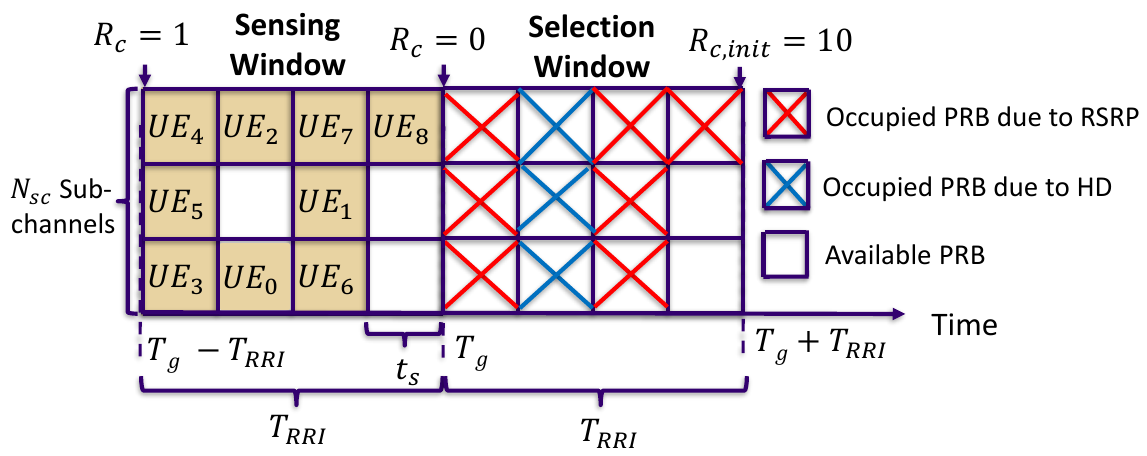}
    \caption{Sensing Window and Selection Window.}
    \label{fig:SenSel}
\end{figure}

 The structure of the Sensing and Selection windows is shown in Fig. \ref{fig:SenSel}. Each UE reserves a PRB for a periodic packet transmission within $T_{RRI}$.  For simplicity, the {\em Sensing window} covers the duration  $[T_{g}-T_{RRI},T_{g}]$ \footnote{Per the 3GPP standard, the sensing Window is defined as the duration $[T_{g}-T_{RRI},T_{g}-T_0]$, where the processing time $T_0$ duration is typically equal to 1 or 2 slots, which is significantly smaller than $T_{RRI}$ duration.}, where $T_{g}$ is the packet generation instant at $R_c = 0$, followed by the {\em Selection window} over the duration $[T_{g}, T_{g} + T_{RRI}]$ \footnote{In such a selection window, the access delay (the elapsed time from the instant packet generates to the instant it transmits) ranges from 1 slot to the maximum packet delay budget, i.e., $T_{RRI}$.}.  If we denote the slot duration as $t_s$ that depends on the numerology, the total number of slots within the sensing/selection window equals $T_{RRI}/t_s$ and consequently, the total number of PRBs within the sensing/selection window is given by 
    \begin{equation}\label{eq:N_r}
       N_r = \frac{T_{RRI}N_{sc}}{t_s}.
    \end{equation}
It should be noted the selection window size can range from 1 slot upto $T_{RRI}$ according to the 3GPP standard. $N_r$ can be readily adapted by replacing $T_{RRI}$ in Eq. (\ref{eq:N_r}) with any selection window size from 1 slot to $T_{RRI}$. In this paper, however, we consider the selection window size to be $T_{RRI}$, which more adequately characterizes the collision events and key SPS parameters with their mutual impacts on the PRR model.

\vspace{-0.8cm}

\subsection{MAC Collision Events in the Selection Window}

MAC collision happens when two (or more) UEs transmit in the same PRB, and all other UEs fail to decode any transmitted packets. Such a MAC collision event starts in a selection window and is followed by consecutive collisions at the next several RRIs. When reference UE 0's $R_C$ decrements to 0, it enters the selection window and either performs reselection with probability $1-p_k$, or does not with probability $p_k$. The MAC collisions in the selection window are thereby conditioned on whether UE 0 performs (or does not perform) reselection in the selection window, i.e.:

\begin{figure}[ht]
    \centering    \includegraphics[width=.55\textwidth]{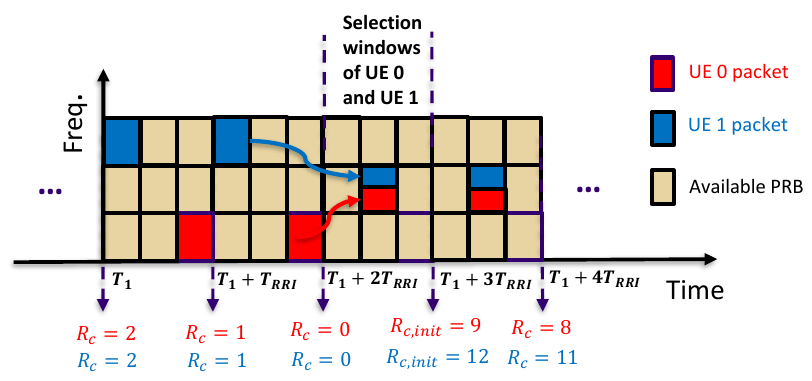}
    \caption{Collision Event 1.}
    \label{fig:ColE1}
    \vspace{-0.2cm}
\end{figure}

 \textbf{Collision Event 1 - UE 0 performs reselection in the selection window and collides with another UE that also performs reselection.} In Fig. \ref{fig:ColE1} we see when UE 0's $R_c$ decrements to 0 at $T = T_1+2T_{RRI}$, UE 1's $R_c$ also decrements to 0, indicating that both enter the selection window simultaneously. Each performs reselection by randomly and independently choosing one of 7 available PRBs out of $N_r = 9$ PRBs in the selection window during [$T_1+2T_{RRI}, T_1+3T_{RRI}$]. If two UEs select the same PRB, it leads to MAC collision in their overlapping selection window. At $T = T_1 + 3T_{RRI}$, $R_{c, init}^{UE 0}(T) = 9$ while $R_{c, init}^{UE 1}(T) = 12$, indicating that the PRB reselected by UE 0 and UE 1 in the selection window will be reserved for the following 9 RRIs and 12 RRIs, respectively. The number of consecutive collisions is thus $\min\left\{R_{c, init}^{UE 0}(T), R_{c, init}^{UE 1}(T) \right \}$. Since both $R_{c, init}^{UE 0}(T)$ and $R_{c, init}^{UE 1}(T) \in [5, 15]$, the number of consecutive collisions right after UE 0's selection window lies in the range [5, 15].

\begin{figure}[t]
\centering
 \subfigure[Sub-Event 1]
{\includegraphics[width=0.45\textwidth]{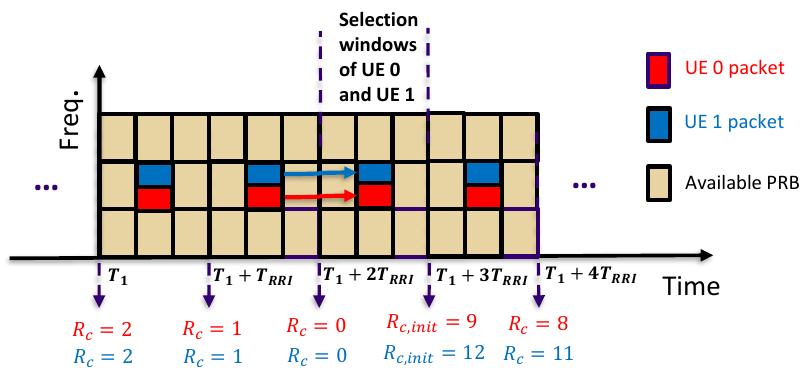}}
\label{fig:6a}
\centering
\subfigure[Sub-Event 2]{
\includegraphics[width=0.45\textwidth]{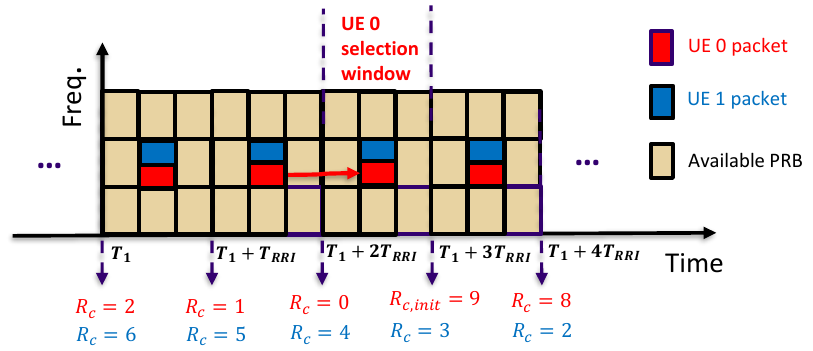}}
\label{fig:6b}
\caption{Collision Event 2.}
\label{fig:ColE2}
\vspace{-0.7cm}
\end{figure}

\textbf{Collision Event 2 - UE 0 does not perform reselection in the selection window and continues to collide with a UE that it has previously collided with in prior RRIs.} Collision Event 2 includes two collision sub-events that depend on the collided UE's $R_c$ state at the moment UE 0's $R_c = 0$: 

\begin{itemize}

\item \textbf{Sub-event 1 - Collided UE's $\bm{R_c = 0}$ when UE 0's $\bm{R_c = 0}$.} In Fig. \ref{fig:ColE2}(a) we see UE 0's $R_c$ decrements to 0 at $T = T_1+2T_{RRI}$, meanwhile the collided UE 1 also decrements to 0. Then, both UEs do not perform reselection, indicating that both persist with the PRB reserved in the prior selection window to the current, i.e., over [$T_1+2T_{RRI}, T_1+3T_{RRI}$]. Since both UEs have collided with each other by reserving the same PRB in the prior RRIs, MAC collisions between UE 0 and UE 1 will continue in their overlapping selection window. The number of consecutive collisions right after UE 0's selection window under Collision Event 2 Sub-event 1 is still $\min\left\{R_{c, init}^{UE 0}(T), R_{c, init}^{UE 1}(T) \right \}$ which lies in the range [5, 15].


\item \textbf{Sub-event 2 - Collided UE's $\bm{R_c \neq 0}$ when UE 0's $\bm{R_c = 0}$.}
In Fig. \ref{fig:ColE2}(b) we see UE 0's $R_c$ decrements to 0 at $T = T_1+2T_{RRI}$, meanwhile the collided UE 1's $R_c = 4$. In the UE 0's selection window during [$T_1+2T_{RRI}, T_1+3T_{RRI}$], UE 0 does not perform reselection by sticking with the PRB reserved in the prior RRIs; meanwhile, UE 1 still uses the PRB reserved in the past RRIs to send packet because UE 1's $R_c$ has not decremented to 0 yet. Since both UEs have collided with each other by reserving the same PRB in the RRIs prior to UE 0's selection window, MAC collisions will continue between UE 0 and UE 1 after UE 0's selection window. The number of consecutive collisions after UE 0's selection window under Collision Event 2 Sub-event 2 is $\min\left\{R_{c, init}^{UE 0}(T), R_{c}^{UE 1}(T) \right \}$. As $R_{c, init}^{UE 0}(T) \in [5, 15]$ and $R_{c}^{UE 1}(T) \in [0, 14]$, the number of consecutive collisions after UE 0's selection window under Collision Event 2 Sub-event 2 lies in the range $[0, 14]$.
\end{itemize}

Finally, note that if UE 0 does not suffer a collision in its selection window, no further collisions occur in the following $R_{c,init}^{UE0}$ RRIs after UE 0's selection window. Since the PRB selected by UE 0 in the selection window is reserved for the following $R_{c,init}^{UE0}$ RRIs, this occupied PRB must be excluded in the following $R_{c,init}^{UE0}$ RRIs for any UE that enters its selection window in this interval.




\section{SPS PRR analytical model}
\label{SPSModel}
  As Sec. \ref{systemSetup} shows, the collision of reference UE 0 in the selection window includes Collision Events 1 and 2. Let $P_{COL}$ be the total probability of MAC collision, $P_{COL,1}$ as the probability for Collision Event 1, and $P_{COL,2}$ as the probability for Collision Event 2. Since Events 1 and 2 are mutually exclusive, 
\begin{equation}\begin{aligned}\label{eq:P_c}
    & P_{COL} = P_{COL,1} + P_{COL,2},
\end{aligned}\end{equation}
where $P_{COL,1}$ and $P_{COL,2}$ are analyzed in the following sections.

\begin{figure*}[htbp]
\centering
\begin{minipage}[t]{0.45\textwidth}
\centering
\includegraphics[width=.98\textwidth]{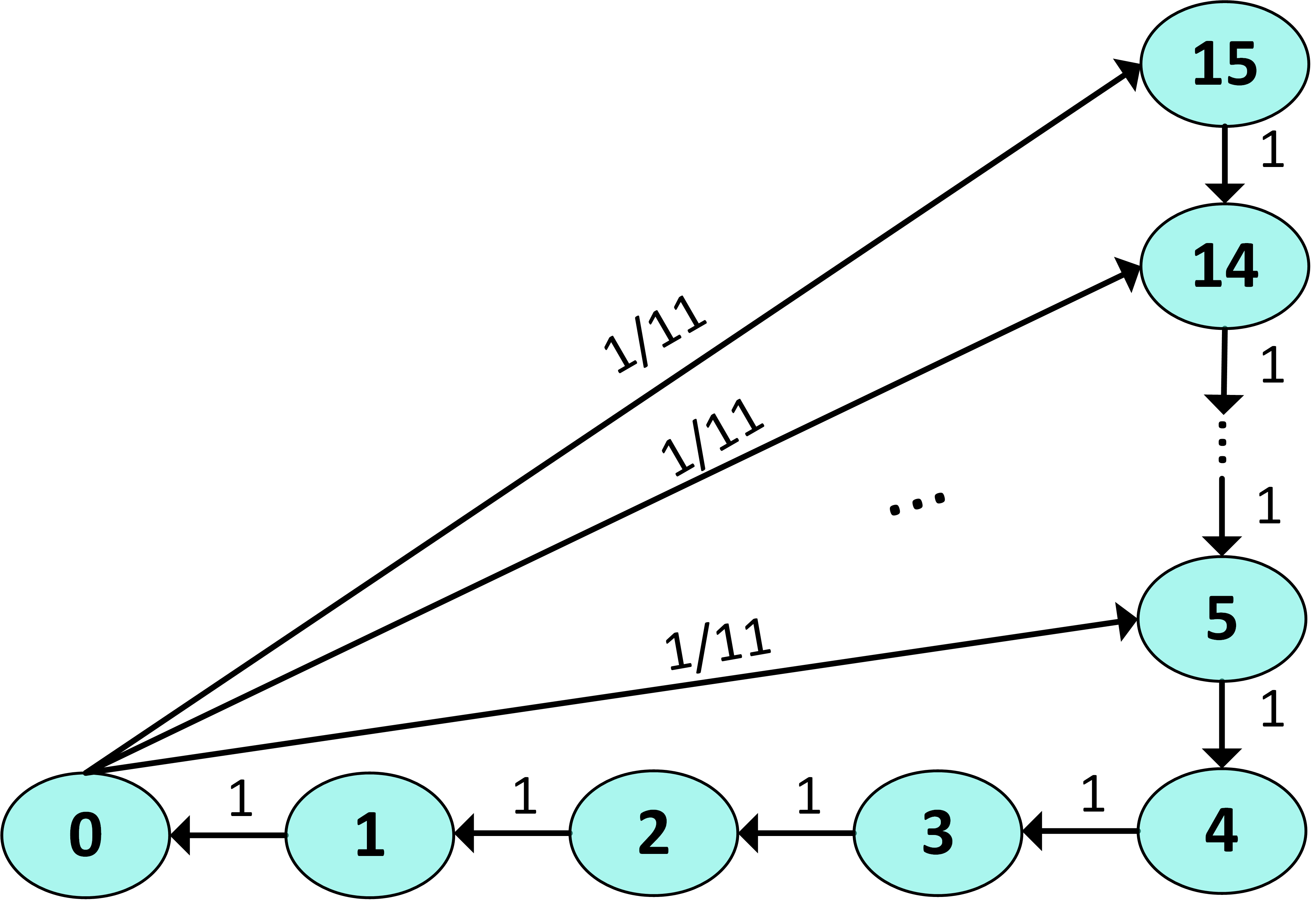}
\caption{Reselection Counter ($R_c$) State Diagram.}
\label{fig:MC_RC}
\end{minipage}
\begin{minipage}[t]{0.45\textwidth}
\centering
\includegraphics[width=.98\textwidth]{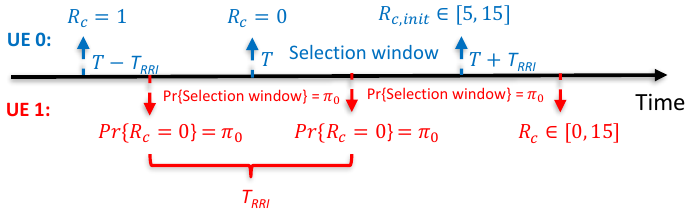}
\caption{Asynchronous $R_c$ Decrement.}
\label{fig:AsynRC}
\end{minipage}
\vspace{-0.5cm}
\end{figure*}


\subsection{Performance Analysis - Reselection Counter State}

Fig. \ref{fig:MC_RC} shows $R_c$ state diagram for our SPS model, identical for all UEs. At any $nT_{RRI}$, the corresponding $R_c(nT_{RRI}) \, \in \, [0, \, 15]$. If $R_c(nT_{RRI}) = 0$, it is randomly re-initialized s.t. $R_{c, init}((n+1)T_{RRI}) \, \in\, [5, 15]$. Thus, the probability that 
    \begin{equation}
      \text{Pr}\{R_{c, init}((n+1)T_{RRI}) = i|R_c(nT_{RRI}) = 0\} = \frac{1}{11}.
    \end{equation}
Denote $\pi_i$ as the probability that $R_c (nT_{RRI}) = i, 0\leq i \leq 15$. According to Fig. \ref{fig:MC_RC}, $\pi_i$ satisfies 


\begin{equation}\label{eq:pi_i}
    \pi_i=
    \begin{cases}
        \pi_0 &  \text{for } 0 \leq i \leq 4,  \\
        \frac{1}{11}\pi_0 + \pi_{i+1} &  \text{for } 5 \leq i \leq 14, \\
        \frac{1}{11}\pi_0 & \text{for } i =  15.
    \end{cases}
\end{equation}

Using the normalization condition $ \sum_{i = 0}^{15}\pi_i = 1$, and solving (\ref{eq:pi_i}), we obtain $\pi_0 = \frac{1}{11}$. Notice that the proposed $R_c$ state diagram in Fig. \ref{fig:MC_RC} can be easily extended to cover different $R_{c,init}$ ranges for $T_{RRI} < 100$ ms, if necessary, which leads to a moderate revision in Eq. (\ref{eq:pi_i}) following a similar calculation manner. Since investigating the impact of the $R_{c,init}$ range is not the focus in this paper, using $R_{c,init} \in [5,15]$ for $T_{RRI} \geq 100$ ms is used for model simplicity.  

We consider asynchronous $R_c$ decrement whereby UEs' $R_c$ counters follow independent clocks. Therefore, UEs' selection windows will partially overlap (when their $R_c$ decrement to 0) within an $T_{RRI}$. Fig. \ref{fig:AsynRC} shows a two-UE example for asynchronous $R_c$ decrement, where UE 0's selection window spans two consecutive RRIs of UE 1, either of which may be UE 1's selection window. Note that since the probability that any UE 1 RRI is its selection window equals $\pi_0$, the probability that UE 0's selection window partially overlaps with UE 1's is $ 2\pi_0$.



\subsection{Performance Analysis - Collision Event 1}

Collision Event 1 happens if at least one of the other UEs also selects the same available PRB as UE 0 (reference UE). Thus, Collision Event 1 is triggered by the reselection behavior of multiple UEs in the overlapping selection window. Suppose UE 0's $R_c = 0$ at time $T$ and UE 0 then performs reselection in its selection window during [$T, T+ T_{RRI}$]. If another UE among $N_{UE}$ UEs participates in reselection during [$T-T_{RRI}, T+ T_{RRI}$], its $R_c$ should first decrement to 0 during [$T-T_{RRI}, T+ T_{RRI}$] (whose probability is $2\pi_0$), and then this UE performs reselection with probability $1-p_k$. As each UE's $R_c$ state is independent, if any $n$ out of total $N_{UE}$ UEs participate in reselection during [$T-T_{RRI}, T+ T_{RRI}$], the corresponding probability follows the binomial distribution, given by
\vspace{-0.3cm}
{\begin{equation}\label{eq:P_r}
\resizebox{.99\hsize}{!}{$
\begin{aligned}
    &\text{Pr\{$n$ UEs $R_c = 0$, Reselection}|\text{$R_c^{UE 0} = 0$, Reselection\}}  = P_s(n) = \begin{pmatrix} N_{UE}\\ n \end{pmatrix} \Bigl(2\pi_0(1-p_k) \Bigl)^n\Bigl(1-2\pi_0(1-p_k)\Bigl)^{N_{UE}-n}.\end{aligned}
    $}
\end{equation}

The UE 0 and $n$ UEs then randomly reselect one of the available (unoccupied) PRBs in the overlapping selection window. The collision occurs if at least one of these $n$ UEs selects the same available (unoccupied) PRB as UE 0. We define such a collision due to $n$-UE participating in reselection in the overlapping selection window as \textbf{$\bm{n}$-fold collision}. Note that the number of available PRBs in the selection window $N_a$ depends on the total number of PRBs in the selection window $N_r$ and the number of occupied PRBs in the selection window $N_o$, i.e., 
\begin{equation}\label{eq:N_a_rv}
    N_a = N_r - N_o,
    \vspace{-0.5cm}
\end{equation}
where both $N_a$ and $N_o$ are random variables. Since each UE with a packet to send needs to select a PRB in each RRI (including the selection window), $N_{UE}$ UEs will occupy $N_{UE}$ PRBs in each RRI if no collision happens (i.e., $N_o = N_{UE}$, a fixed number). However, in the event of collisions, $N_o < N_{UE}$ is a random variable. The average number 
 of occupied PRBs $\overline{N_o}$ in the selection window can be expressed in terms of the total MAC collision probability,
\begin{equation}\label{eq:N_o}
    \overline{N_o} = N_{UE} - P_{COL} N_{UE} +\frac{P_{COL} N_{UE}}{\overline{N_c}},
\end{equation}
where $P_{COL}$ is the total MAC collision probability (Collision Event 1 + Collision Event 2) to be determined, and the random variable $N_c \geq 2$ is the number of UEs in an occupied PRB where the collision happens. In Eq. (\ref{eq:N_o}), $P_{COL} N_{UE}$ represents the average number of UEs who collide with other UEs in the selection window. Thus $\frac{P_{COL} N_{UE}}{\overline{N_c}}$ represents the average number of occupied PRBs where collisions happen in the selection window, while $N_{UE} - P_{COL} N_{UE}$ represents the average number of occupied PRBs where collisions do not happen in the selection window. As Eq. (\ref{eq:N_o}) suggests, $\overline{N_o}$ will decrease if $\overline{N_c}$ increases. Then the average number of available (unoccupied) PRBs in the selection window, $\overline{N_a}$, is given by
\begin{equation}\label{eq:N_a}
    \overline{N_a} = N_r - \overline{N_o} = N_r - N_{UE} + \frac{(\overline{N_c}-1)P_{COL} N_{UE}}{\overline{N_c}}.
\end{equation}


Given $\overline{N_a}$ available (unoccupied) PRBs in the selection window, the probability that one of $n$ UEs collides with UE 0 by reselecting the same available PRB is $1/\overline{N_a}$. As a result, the probability that $n$-fold collision happens (i.e., at least one of $n$ UEs select the same PRB as UE 0) given that $n$ UEs participate in reselection is \cite{wei2024optimized}
\begin{equation}\begin{aligned}\label{eq:P_s_1}
    \text{Pr\{$n$-fold Collision}|\text{$n$ UEs $R_c = 0$, Reselection\}} = P_r(n) = 1- \left(1-\frac{1}{\overline{N_a}}\right)^{n}.
\end{aligned}\end{equation}
Therefore, the probability that $n$-fold collision happens given that UE 0 performs reselection in the selection window is $P_r(n)P_s(n)$. Considering all  $n$ $(1\leq n \leq N_{UE})$, we can obtain the collision probability when UE 0 performs reselection in the selection window as: 
\begin{equation}\begin{aligned}\label{eq:P_c^sr_1}
    \text{Pr}\{\text{Collision}|\text{$R_c^{UE 0} = 0$, Reselection\}} =\sum_{n=1}^{N_{UE}}P_r(n)P_s(n).\end{aligned}
\end{equation}
Substituting Eq. (\ref{eq:P_r}) and (\ref{eq:P_s_1}) into Eq. (\ref{eq:P_c^sr_1}), yields 

\begin{equation}\begin{aligned}\label{eq:P_c1}
    \text{Pr}\{\text{Collision}|\text{$R_c^{UE 0} = 0$, Reselection\}}  = 1- \left( 1 - \frac{2\pi_{0}(1-p_k)}{\overline{N_a}} \right)^{N_{UE}}.\end{aligned}
\end{equation}

$P_{COL,1}$ is thus expressed as

\begin{equation}
\begin{aligned}
\label{eq:P_c1_complete}
   P_{COL,1} &= \text{Pr}\{\text{UE 0 Reselection}|\text{$R_c^{UE 0} = 0$\}}\cdot\text{Pr}\{\text{Collision}|\text{$R_c^{UE 0} = 0$, Reselection\}} \\&=(1-p_k)\left[1- \left( 1 - \frac{2\pi_{0}(1-p_k)}{\overline{N_a}} \right)^{N_{UE}}\right],
\end{aligned}
\end{equation}
where $\overline{N_a}$ is given by Eq. (\ref{eq:N_a}). Note that for $p_k = 0$, $P_{COL} = P_{COL,1}$ from Eq. (\ref{eq:P_c}) and hence $P_{COL,1}$ can be directly determined by solving Eq. (\ref{eq:N_a}) and (\ref{eq:P_c1_complete}) simultaneously.
However for $p_k > 0$, $P_{COL,1}$ cannot be directly determined yet because $P_{COL,1}$ is a function of $P_{COL,2}$. Meanwhile, according to Eq. (\ref{eq:P_c1_complete}), $P_{COL,1}$ decreases as $\overline{N_a}$ increases, and hence from Eq. (\ref{eq:N_a}), $P_{COL,1}$ will decrease as $\overline{N_c}$ increases.

\subsection{Performance Analysis - Collision Event 2}

\begin{figure}[t]
    \centering
\includegraphics[width=0.55\linewidth]{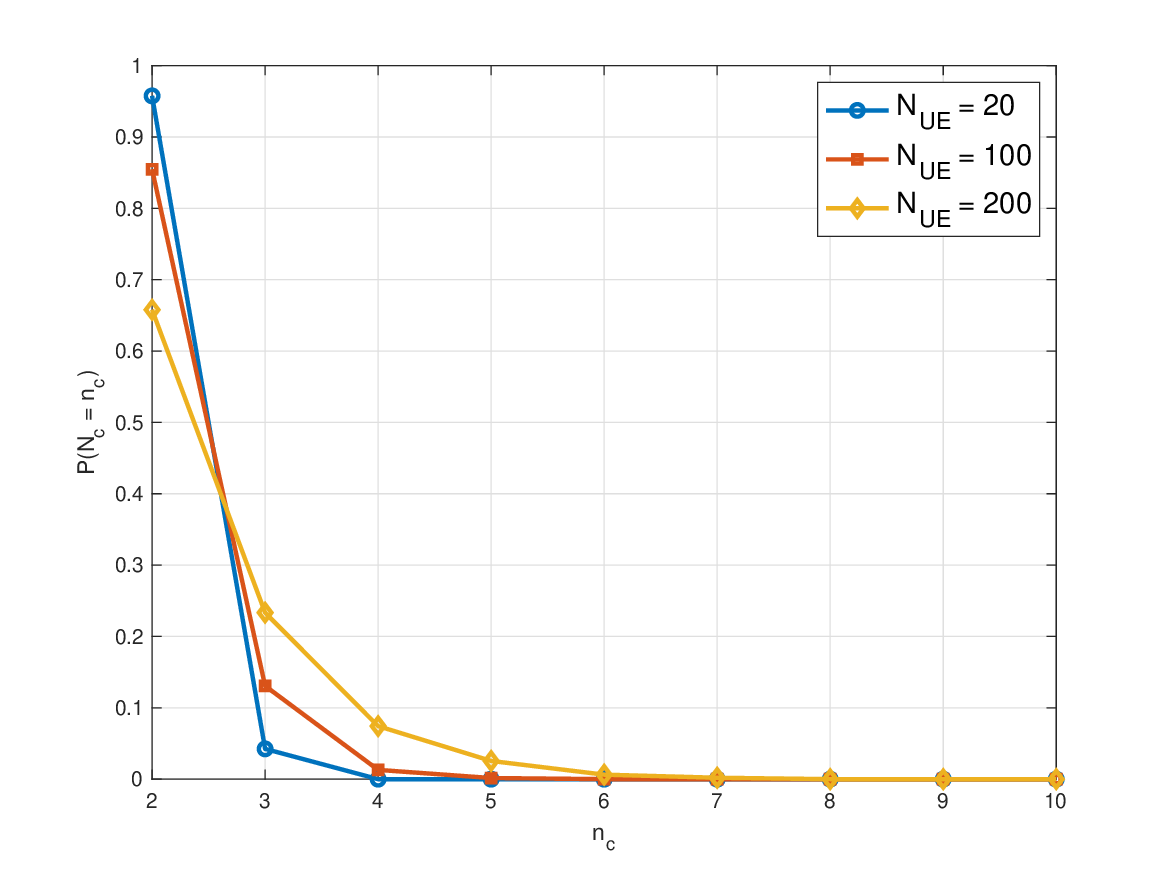}    
\caption{The Proportion of $n_c$-Packet Collision at Different $N_{UE}$.}
    \label{fig:Pnc}
    \vspace{-0.5cm}
\end{figure}

Collision Event 2 happens if UE 0 has collided with at least one of the other UEs in UE 0's most recent selection window, and collision will continue in UE 0's current selection window if both UE 0 and the collided UE(s) do not change the PRB after UE 0's most recent selection window. 
Similar to Collision Event 1 potentially involving multiple ($ >2 $) packets/UEs as indicated by Eq. (\ref{eq:P_s_1}) resulting in $n$-fold collision, Collision Event 2 may also involve multiple packets/UEs. For instance, if $n = 10$, ten other UEs participate in reselection within the overlapping selection window as UE 0, the number of collided packets/UEs in a collision $N_c$ can be up to 10. However, Fig. \ref{fig:Pnc} shows that the 2-packet/UE collision (i.e., $N_c = 2$) probability dominates for all $N_{UE}$, as observed from data gathered through ns-3 simulations, especially in under-saturated condition. Since our analytical model is aimed at the under-saturated condition, we hereafter assume that only the 2-packet/UE collisions constitute Collision Event 2 to simplify the analytical model, as shown in Fig. \ref{fig:ce22sub}. 

Since Collision Event 2 occurs conditioned on the events in UE 0's {\em most recent} previous selection window during [$T_1, T_1 + T_{RRI}$], as represented by the following event sets: 




\begin{figure}[t]
\centering
 \subfigure[Sub-event 1: E1 Followed by no Reselection of both UEs.]
{\includegraphics[width=0.65\textwidth]{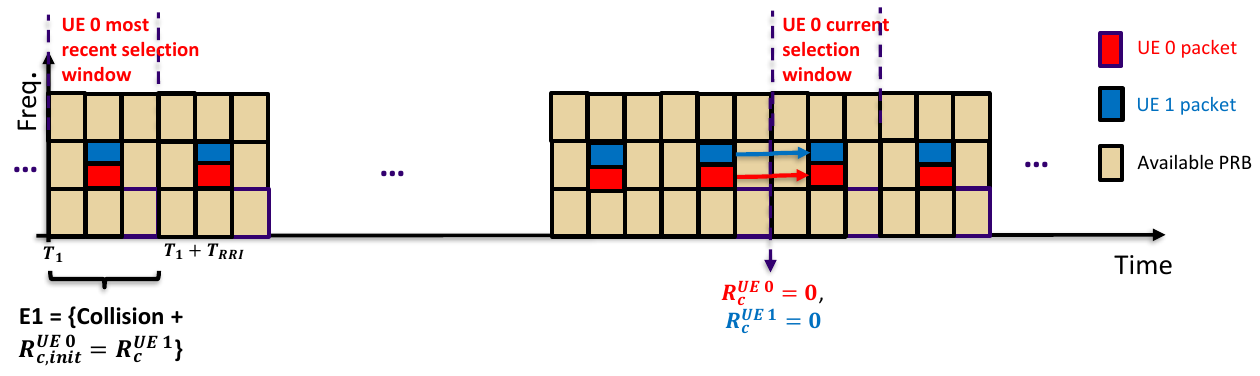}}
\label{fig:13a}
\centering
\subfigure[Sub-event 2: E2 Followed by no Reselection of UE 0.]{
\includegraphics[width=0.65\textwidth]{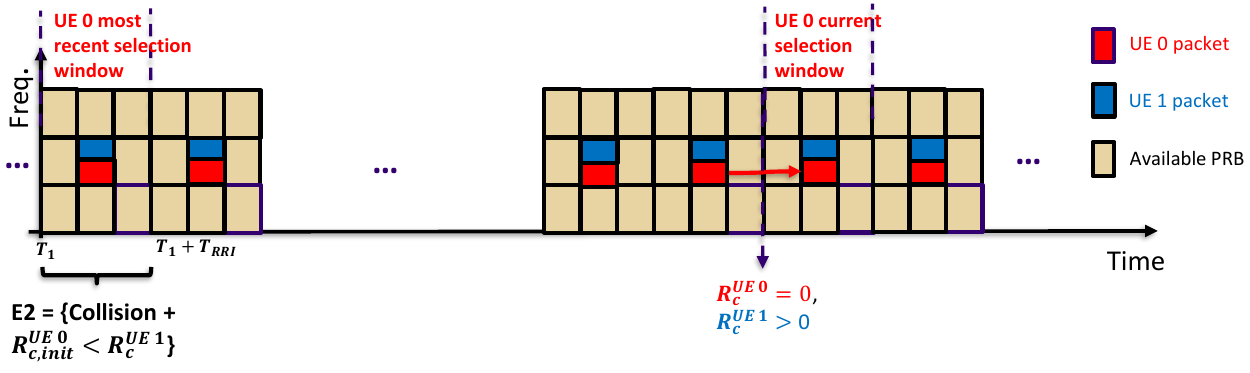}}
\label{fig:13b}
\centering
\subfigure[Sub-event 2: E3 Followed by no Reselection of both UEs.]{
\includegraphics[width=0.65\textwidth]{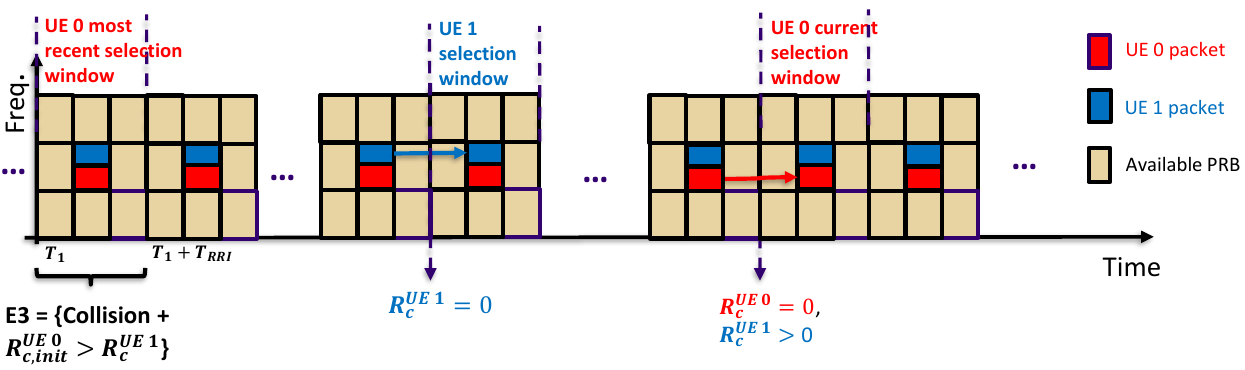}}
\label{fig:13c}
\caption{Events During [$T_1, T_1 + T_{RRI}$] Followed by UEs' Behaviors that leading Collision Event 2.}
\label{fig:ce22sub}
\vspace{-0.5cm}
\end{figure}

\begin{itemize}
    \item $E 1 = \Bigl\{ \text{Collision during } [T_1, T_1 + T_{RRI}], \text{and } R_{c,init}^{UE 0} = R_{c}^{UE 1} \text{ at } T = T_1 + T_{RRI} \Bigl\}$, shown in Fig. \ref{fig:ce22sub}(a). Then $R_c^{UE 0} = R_c^{UE 1} = 0$ at the beginning of UE 0's current selection window, and collision will continue into UE 0's current selection window if both UEs do not perform reselection (with probability = $p_k^2$). Since the probability that $E_1$ happens is given by:
    \begin{equation}\begin{aligned}\label{eq:P_c21_c}
    &\text{Pr}\{E1\} = P_{COL} \cdot \text{Pr}\{R_{c,init}^{UE 0} = R_{c}^{UE 1}\} ,\end{aligned}
\end{equation}
where $\text{Pr}\{R_{c,init}^{UE 0} = R_{c}^{UE 1}\} = \text{Pr}\{R_c^{UE 1} = 0|R_c^{UE 0} = 0\}= 2\pi_0$. Then, the probability that Collision Event 2 happens due to $E1$ is thereby derived as
    \begin{equation}\begin{aligned}\label{eq:P_c21}
    &P_{COL,2}^{E1} = \text{Pr}\{\text{Collision}|E1\} \cdot\text{Pr}\{E1\}.\end{aligned}
\end{equation}
 where $\text{Pr}\{\text{Collision}|E1\} = p_k^2$.

    \item $E 2 = \Bigl\{\text{Collision during } [T_1, T_1 + T_{RRI}], \text{and } R_{c,init}^{UE 0} < R_{c}^{UE 1} \text{ at } T = T_1 + T_{RRI}\Bigl\}$, shown in Fig. \ref{fig:ce22sub}(b). Then $R_c^{UE 0} = 0$ while $R_c^{UE 1} > 0$ at the beginning of UE 0's current selection window, and collision will continue in the current reselection window if UE 0 does not perform reselection (with probability = $p_k$). The probability that $E_2$ happens is given by:
\begin{equation}\label{eq:P_c22_c}
    \resizebox{.8\hsize}{!}{$
    \begin{aligned}
        &\text{Pr}\{E2\} = P_{COL} \cdot \text{Pr}\{R_{c,init}^{UE 0} < R_{c}^{UE 1}\} =  P_{COL}\cdot\sum_{|I|}\Bigl[\text{Pr}\{ R_{c, init}^{UE0} = i \}\cdot \text{Pr}\{R_{c, init}^{UE0}< R_{c}^{UE1}| R_{c, init}^{UE0} = i \}\Bigl],\end{aligned}
        $}
    \end{equation}
    
where $i \in I = \{5, 6, ..., 15\}$ which is the set of all $R_{c,init}$ states, and $\text{Pr}\{R_{c,init}^{UE 0} < R_{c}^{UE 1}\}$ can be obtained according to Fig. \ref{fig:MC_RC}. Then, the probability that Collision Event 2 happens due to $E2$ is thereby derived as

    \begin{equation}\label{eq:P_c22}
    \begin{aligned}
        &P_{COL,2}^{E2}  = \text{Pr}\{\text{Collision}|E2\} \cdot\text{Pr}\{E2\},\end{aligned}
    \end{equation}
 where $\text{Pr}\{\text{Collision}|E2\} = p_k$.

    \item $E 3 = \Bigl\{\text{Collision during } [T_1, T_1 + T_{RRI}], \text{and } R_{c,init}^{UE 0} > R_{c}^{UE 1} \text{ at } T = T_1 + T_{RRI}\Bigl\}$, shown in Fig. \ref{fig:ce22sub}(c). Then $R_c^{UE 1} = 0$ and UE 1 does not perform reselection\footnote{We ignore the second-order events and always assume that $R_{c,init}^{UE 1} > R_{c}^{UE 0}$ at the end of UE 1's selection window.} (with probability = $p_k$) before UE 0's current selection window. Afterward, $R_c^{UE 0} = 0$ while $R_c^{UE 1} > 0$ at the beginning of UE 0's current selection window, and the collision will continue if UE 0 does not perform reselection (with probability = $p_k$). Then the probability that $E_3$ happens is given by:
        \begin{equation}\label{eq:P_c23_c}
    \resizebox{.85\hsize}{!}{$
    \begin{aligned}
    &\text{Pr}\{E3\} = P_{COL} \cdot \text{Pr}\{R_{c,init}^{UE 0} > R_{c}^{UE 1}\} =  P_{COL}\cdot\sum_{|I|}\Bigl[\text{Pr}\{ R_{c, init}^{UE0} = i \}\cdot \text{Pr}\{R_{c, init}^{UE0} > R_{c}^{UE1}| R_{c, init}^{UE0} = i \}\Bigl],\end{aligned}
    $}
\end{equation}

where $i \in I = \{5, 6, ..., 15\}$, and $\text{Pr}\{R_{c,init}^{UE 0} > R_{c}^{UE 1}\}$ can be obtained according to Fig. \ref{fig:MC_RC}. The probability that Collision Event 2 happens due to $E3$ is thereby derived as
    \begin{equation}\begin{aligned}\label{eq:P_c23}
    &P_{COL,2}^{E3} = \text{Pr}\{\text{Collision}|E3\} \cdot\text{Pr}\{E3\},\end{aligned}
\end{equation}
where $\text{Pr}\{\text{Collision}|E3\} = p_k^2$.

\end{itemize}

As $P_{COL,2}^{E1}$, $ P_{COL,2}^{E2}$ and $ P_{COL,2}^{E3}$ are \textbf{mutually exclusive}, $P_{COL,2}$ can be thereby finalized as:
\begin{equation}
\begin{aligned}
\label{eq:P_c2_complete}
&   P_{COL,2} = P_{COL,2}^{E1} + P_{COL,2}^{E2} + P_{COL,2}^{E3}
.
\end{aligned}
\end{equation}




\subsection{Performance Analysis - Packet Reception}

The closed-form of MAC collision probability $P_{COL}$, is obtained by substituting Eq. (\ref{eq:P_c1_complete}) and (\ref{eq:P_c2_complete}) into (\ref{eq:P_c}), where $P_{COL}$ can be numerically calculated by solving Eq. (\ref{eq:N_a}) and (\ref{eq:P_ctot}) simultaneously.

\begin{equation}\label{eq:P_ctot}
\resizebox{.6\hsize}{!}{$
   P_{COL} = \frac{(1-p_k)\left(1- \left( 1 - \frac{2\pi_{0}(1-p_k)}{\overline{N_a}} \right)^{N_{UE}}\right)}{1-\Big( 2\pi_0p_k^2 + 
p_k \text{Pr}\{R_{c,init}^{UE 0} < R_{c}^{UE 1}\} + p_k^2\text{Pr}\{R_{c,init}^{UE 0} > R_{c}^{UE 1}\}\Big) },
   $}
\end{equation}
where the calculation of $\text{Pr}\{R_{c,init}^{UE 0} < R_{c}^{UE 1}\}$ and $\text{Pr}\{R_{c,init}^{UE 0} > R_{c}^{UE 1}\}$ are shown in Appendix \ref{cal1} and \ref{cal2}, respectively. We further derive $PRR$ - the probability that a packet is received successfully - which is determined by 
\begin{itemize}
    \item \emph{MAC collision errors}, i.e., Two (or more) UEs transmit packets in the same PRB, and all receive UEs fail to decode those packets; this occurs with probability $P_{COL}$;
    \item \emph{Half-duplex (HD) errors}. These occur because UEs are assumed to be half-duplex devices, incapable of transmitting and receiving simultaneously. Thus, if a transmitter UE and a receiver UE transmit their packets in the same slot, the receiver UE cannot decode the packet from the transmitting UE. The probability that HD errors happen depends on the number of slots in each RRI, given by $P_{HD} = \frac{t_s}{T_{RRI}}$.
 
\end{itemize} 
    If a packet is successfully received, neither MAC collision errors nor HD errors must occur. Thus $PRR$ is given by 
\begin{equation}\label{eq:Prr_single}
   PRR =  (1-P_{COL})(1-P_{HD}).
\end{equation}



\section{Model Extension}
\label{extension}

\subsection{Multiple Layer-2 Packet Transmissions per RRI}
To improve the reliability of SL communication, the Packet Data Convergence Protocol (PDCP) duplication mechanism has been standardized in 3GPP \cite{3gpp.37.985}, whereby a PDCP Packet Data Unit (PDU) is duplicated into two (or more) instances and transmitted on two (or more) PRBs. In this section, we investigate the impact of multiple Layer-2 packet transmissions per RRI.

Since each UE now uses $N_{Se}$ PRBs to transmit $N_{Se}$ identical copies of the PDCP PDU in each RRI, $N_{UE}$ UEs will require $N_{UE}N_{Se}$ PRBs in each RRI as the initial resource pool. Following the logic leading to Eq. (\ref{eq:N_o}), the resulting average number of occupied PRBs in the selection window $\overline{N_o}(N_{Se})$, becomes
\begin{equation}\label{eq:N_o_extension}
\resizebox{.6\hsize}{!}{$
\begin{aligned}
    &\overline{N_o}(N_{Se}) =  N_{UE}N_{Se} - P_{COL}(N_{Se}) N_{UE}N_{Se} +\frac{P_{COL}(N_{Se}) N_{UE}N_{Se}}{\overline{N_c}(N_{Se})}.
\end{aligned}
$}
\end{equation}

Then the average number of available (unoccupied) PRBs in the selection window, $\overline{N_a}(N_{Se})$, is given by

\begin{equation}\label{eq:N_a_extension}
\resizebox{.8\hsize}{!}{$
\begin{aligned}
   &\overline{N_a}(N_{Se}) = N_r - \overline{N_o}(N_{Se}) = N_r - N_{UE}N_{Se} + \frac{(\overline{N_c}(N_{Se})-1)P_{COL}(N_{Se}) N_{UE}N_{Se}}{\overline{N_c}(N_{Se})}.
\end{aligned}
$}
\end{equation}

\begin{figure*}\begin{equation}
\resizebox{.9\hsize}{!}{$
\begin{aligned}
\label{eq:P_c1_complete_X}
   &P_{COL,1}(N_{Se}, X) =(1-p_k)\left\{ \left[1- \left( 1 + 2\pi_0(1-p_k)\left( \left(1-\frac{1}{\overline{N_a}(N_{Se})}\right)^{N_{Se}}-1\right) \right)^{N_{UE}}\right]\cdot\frac{\overline{N_a}(N_{Se})}{XN_r} + 1 \cdot \frac{XN_r - \overline{N_a}(N_{Se})}{XN_r}\right\},
\end{aligned}
$}
\end{equation}
\end{figure*}

\begin{figure*}
    \begin{equation}\label{eq:P_c1_complete_X_overall}
        \resizebox{.95\hsize}{!}{$
            P_{COL}(N_{Se}, X)  =
            \begin{cases}
                P_{COL}(N_{Se}) & \text{for } \overline{N_a}(N_{Se}) \geq XN_r,\\
                \frac{(1-p_k)\left\{ \left[1- \left( 1 + 2\pi_0(1-p_k)\left( \left(1-\frac{1}{\overline{N_a}(N_{Se})}\right)^{N_{Se}}-1\right) \right)^{N_{UE}}\right]\cdot\frac{\overline{N_a}(N_{Se})}{XN_r} + 1 \cdot \frac{XN_r - \overline{N_a}(N_{Se})}{XN_r}\right\}}{1-\Big( 2\pi_0p_k^2 + 
p_k \text{Pr}\{R_{c,init}^{UE 0} < R_{c}^{UE 1}\} +  p_k^2\text{Pr}\{R_{c,init}^{UE 0} > R_{c}^{UE 1}\}\Big)  } & \text{for } \overline{N_a}(N_{Se}) < XN_r.
            \end{cases}
        $}
    \end{equation}
\end{figure*}

In Collision Event 1, the probability that $n$-fold collision happens (i.e., at least one of $nN_{Se}$ packets occupy the same PRB as UE 0) given that $n$ UEs participate in reselection is
\begin{equation}\label{eq:P_s_1_extension}
 P_r(n, N_{se}) = 1- \left(1-\frac{1}{\overline{N_a}(N_{Se})}\right)^{nN_{Se}}.
\end{equation}
Substitute Eq. (\ref{eq:P_r}) and (\ref{eq:P_s_1_extension}) into (\ref{eq:P_c^sr_1}), yielding:
\begin{equation}\begin{aligned}\label{eq:P_c1_extension}
    &\text{Pr}\{\text{Collision}|\text{UE 0 $R_c = 0$, Reselection}\}  = 1- \left( 1 + 2\pi_0(1-p_k)\left( \left(1-\frac{1}{\overline{N_a}}\right)^{N_{Se}}-1\right) \right)^{N_{UE}}.\end{aligned}
\end{equation}

From Eq. (\ref{eq:P_c}) and (\ref{eq:P_c1_extension}), $P_{COL,1}$ in terms of $N_{Se}$ is expressed as

\begin{equation}
\resizebox{.7\hsize}{!}{$
\begin{aligned}
\label{eq:P_c1_complete_extension}
   &P_{COL,1}(N_{Se}) = (1-p_k)\left[1- \left( 1 + 2\pi_0(1-p_k)\left( \left(1-\frac{1}{\overline{N_a}(N_{Se})}\right)^{N_{Se}}-1\right) \right)^{N_{UE}}\right],
\end{aligned}
$}
\end{equation}
where $\overline{N_a}(N_{Se})$ is expressed in Eq. (\ref{eq:N_a_extension}).
Accordingly, $P_{COL,2}$ expressed in Eq. (\ref{eq:P_c2_complete}) is modified as follows 
\begin{equation}
\resizebox{.8\hsize}{!}{$
\begin{aligned}
\label{eq:P_c2_complete_asyn_Nse}
&   P_{COL,2}(N_{Se}) = P_{COL}(N_{Se})\Bigl\{2\pi_0p_k^2 + 
p_k \text{Pr}\{R_{c,init}^{UE 0} < R_{c}^{UE 1}\} +  p_k^2\text{Pr}\{R_{c,init}^{UE 0} > R_{c}^{UE 1}\}\Bigl\}.
\end{aligned}
$}
\end{equation}
Then, the closed form of $P_{COL}(N_{Se})$ can be obtained by substituting Eq. (\ref{eq:P_c1_complete_extension}) and (\ref{eq:P_c2_complete_asyn_Nse}) into Eq. (\ref{eq:P_c}):

\begin{equation}
\resizebox{.7\hsize}{!}{$
\begin{aligned}\label{eq:P_ctot_extension}
   &P_{COL}(N_{Se}) = \frac{(1-p_k)\left[1- \left( 1 + 2\pi_0(1-p_k)\left( \left(1-\frac{1}{\overline{N_a}(N_{Se})}\right)^{N_{Se}}-1\right) \right)^{N_{UE}}\right]}{1-\Big( 2\pi_0p_k^2 + 
p_k \text{Pr}\{R_{c,init}^{UE 0} < R_{c}^{UE 1}\} + p_k^2\text{Pr}\{R_{c,init}^{UE 0} > R_{c}^{UE 1}\}\Big) }.\end{aligned}
   $}
\end{equation}

When $N_{Se} > 1$, a packet transmission is successful if at least one of $N_{Se}$ duplicate packets is correctly decoded. Thus $PRR$ in terms of $N_{Se}$ is given by

\begin{equation}\label{eq:Prr_multi}
\resizebox{.5\hsize}{!}{$
   PRR(N_{Se}) = 1 - \Bigl[1 - \Bigl(1-P_{COL}(N_{Se})\Bigl)(1-P_{HD})\Bigl]^{N_{Se}}.
   $}
\end{equation}
where for $N_{Se} = 1$, $P_{COL}(1)$ in Eq. (\ref{eq:P_ctot_extension}) and $PRR(1)$ in Eq. (\ref{eq:Prr_multi}) defaults to $P_{COL}$ in Eq. (\ref{eq:P_ctot}) and $PRR$ in Eq. (\ref{eq:Prr_single}), as expected.

\subsection{Proportion of available PRBs for reselection}
Another option that possibly improves the SL communication reliability is increasing the minimum proportion of available PRBs for reselection characterized by the parameter $X$, whose range has been standardized in 3GPP \cite{3gpp.36.213, 3gpp.36.321}. In the previous sections, the proportion of available (unoccupied) PRBs for reselection was fixed at the lower limit, $X = 0.2$. In this section, we investigate the impact of (higher) $X$ on $P_{COL}(X, N_{Se})$. We show that $P_{COL}(X, N_{Se})$ does not benefit from increasing $X$, although the number of available PRBs for reselection increases.

In the selection window, the number of available PRBs used for reselection must be first greater than a threshold (i.e., $XN_r$), which is the required minimum number of PRBs. UE randomly reselects among available PRBs for packet transmissions. In the previous sections, since $X$ is set to a small value (e.g., $X = 0.2$), $\overline{N_a}(N_{Se})$, that represents the number of unoccupied PRBs (referred to as the available PRBs in the previous sections), satisfies $\overline{N_a}(N_{Se}) > XN_r$ in the under-saturated condition. However, $\overline{N_a}(N_{Se}) > XN_r$ may not hold if $X$ is increased significantly: UE has to increase $\overline{N_a}(N_{Se})$ to $XN_r$ by including $XN_r - \overline{N_a}(N_{Se})$ occupied PRBs in order to satisfy the minimum number requirement of total $XN_r$ available PRBs for reselection. Thus, the $XN_r$ available PRBs now include $\overline{N_a}(N_{Se})$ unoccupied PRBs and $XN_r - \overline{N_a}(N_{Se})$ occupied PRBs. In Collision Event 1, if $\overline{N_a}(N_{Se}) < XN_r$, two collision cases happen given that UE 0 performs reselection in the selection window: 
\begin{itemize}
    \item \emph{Case 1}: UE 0 selects one of the $\overline{N_a}(N_{Se})$ unoccupied PRBs among $XN_r$ available PRBs, and a collision happens if another UE reselects the same unoccupied PRB as UE 0;
    \item \emph{Case 2}: Collision happens if UE 0 selects one of the $XN_r - \overline{N_a}(N_{Se})$ occupied PRBs among $XN_r$ available PRBs.
\end{itemize}

Regarding Case 1, the probability that UE 0 selects one of the unoccupied PRBs among $XN_r$ available PRBs is $\frac{\overline{N_a}(N_{Se})}{XN_r}$. Meanwhile, the probability that collision happens is expressed in Eq. (\ref{eq:P_c1_extension}); Regarding Case 2, the probability that UE 0 selects one of the occupied PRBs among $XN_r$ PRBs is $\frac{XN_r - \overline{N_a}(N_{Se})}{XN_r}$, and the probability that collision happens is 1. Thus $P_{COL, 1} (X, N_{Se})$ under $\overline{N_a}(N_{Se}) < XN_r$ is given by Eq. (\ref{eq:P_c1_complete_X}), where $\overline{N_a}(N_{Se})$ (that denotes the number of unoccupied PRBs in the selection window) is expressed in Eq. (\ref{eq:N_a_extension}). Then, the closed form of $P_{COL}(N_{Se},X)$ under $\overline{N_a}(N_{Se}) < XN_r$ can be obtained by substituting Eq. (\ref{eq:P_c1_complete_X}) and (\ref{eq:P_c2_complete_asyn_Nse}) into Eq. (\ref{eq:P_c}); meanwhile, $P_{COL}(N_{Se},X) = P_{COL}(N_{Se})$ under $\overline{N_a}(N_{Se}) \geq XN_r$, where $P_{COL}(N_{Se})$ is given by Eq. (\ref{eq:P_ctot_extension}). As a result, $P_{COL}(N_{Se},X)$ is expressed in Eq. (\ref{eq:P_c1_complete_X_overall}), where $\overline{N_a}(N_{Se})$ is expressed in Eq. (\ref{eq:N_a_extension}). Accordingly, the PRR in terms of $X$ is given by
\begin{equation}\label{eq:Prr_multi_X}
\resizebox{.6\hsize}{!}{$
   PRR(N_{Se},X) = 1 - \Bigl[1 - \Bigl(1-P_{COL}(N_{Se},X)\Bigl)(1-P_{HD})\Bigl]^{N_{Se}}.
   $}
\end{equation}
Meanwhile, the average number of available PRBs in the selection window, $\overline{N_a}(N_{Se},X)$, is given by
\begin{equation}\label{eq:NaNseX}
    \overline{N_a}(N_{Se},X) =
    \begin{cases}
        \overline{N_a}(N_{Se}) &  \text{for } \overline{N_a}(N_{Se}) \geq XN_r,  \\
        XN_r &  \text{for } \overline{N_a}(N_{Se}) < XN_r,
    \end{cases}
\end{equation}
where $\overline{N_a}(N_{Se})$ is given by Eq. (\ref{eq:N_a_extension}).

\section{Model Validation}
\label{validation}

\subsection{Simulation Description}
To validate the predicted performance for the asynchronous $R_c$ decrementation based on our model, we resort to simulations using ns-3\footnote{Open source network simulator, available @ www.nsnam.org.} by building off the 5G-LENA module as a base \cite{ali20213gpp,brady2022modeling}\footnote{All code used to generate both modeled and simulated curves can be found at https://github.com/CollinBrady1993/Code-for-NR-C-V2X-Tutorial.}. The main parameters used during the ns-3 simulation can be found in Table \ref{tab:simParams}. 
\begin{table}[ht]
 \centering
 \caption{\small{Main ns-3 Simulation Parameters} 
}\label{tab:simParams}
\resizebox{.35\textwidth}{!}{\begin{tabular}{ |c|c|c|c|c|c|c| } 
\hline
\textbf{Parameter} & \textbf{Value} & \textbf{Parameter} & \textbf{Value} \\
\hline
$N_{sc}$  & 2 &  $P_t$  & 23 dBm\\
\hline
$t_s$  & 1 ms&  $T_{RRI}$  & 100 ms\\
\hline
$\gamma_{SPS}$ & -110 dBm & $N_{UE}$ & [20, 200] \\
\hline
$p_k$& [0, 0.8]& $X$ & [0.2, 0.5] \\
\hline
\end{tabular}}
\end{table}


\begin{figure}[ht]
    \centering
\includegraphics[width=0.7\linewidth]{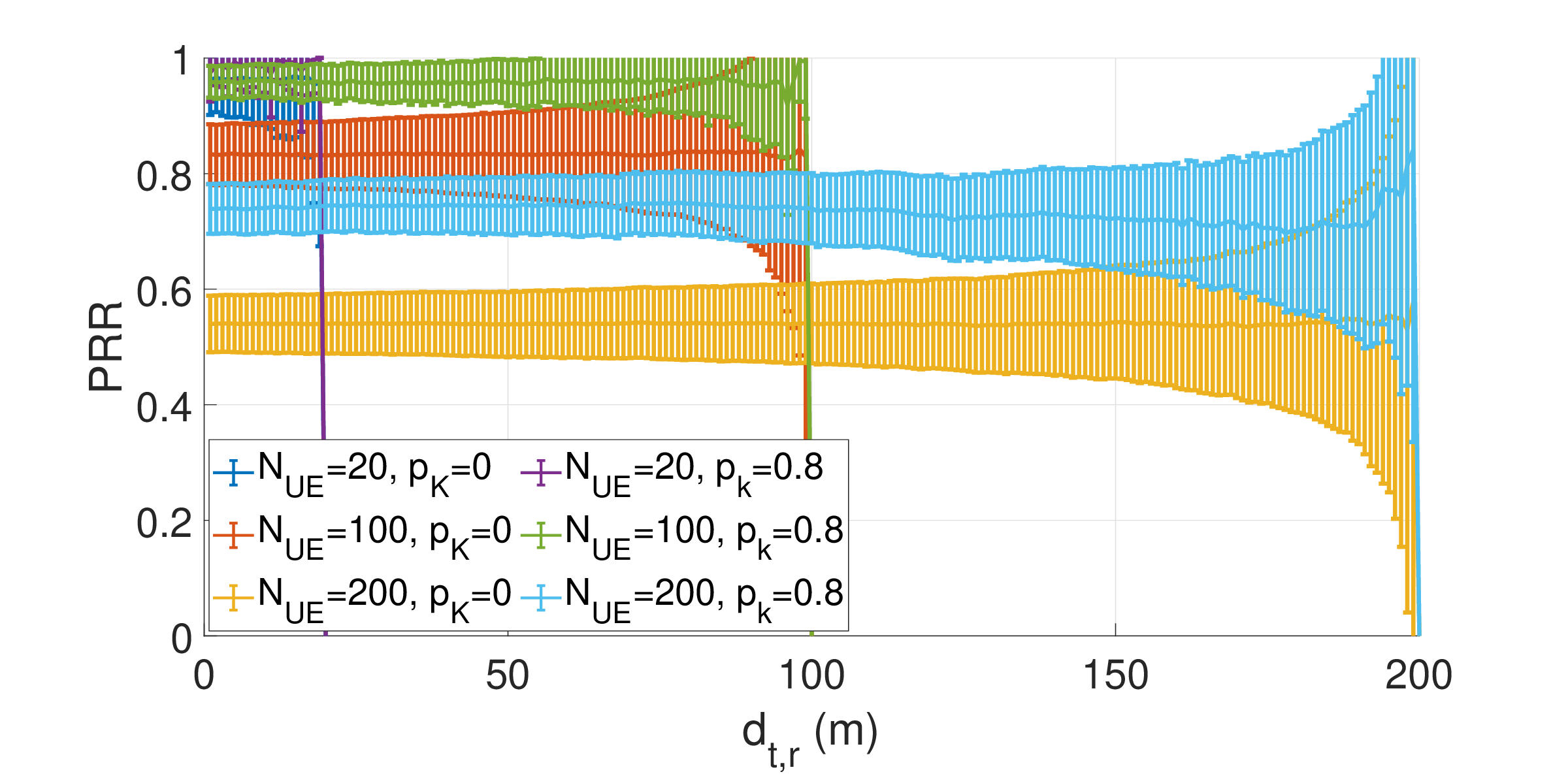}    \caption{Simulated $PRR (d_{t, r})$ in the Fully Connected Network.}
    \label{fig:PRRdependance}
    \vspace{-0.3cm}
\end{figure}

Simulations were conducted for a linear topology with $N_{UE}$ UEs in a single row separated by 1 meter, allowing them to transmit periodic safety messages according to the V2X mode 2 algorithm. Each simulation took place over the course of 100 seconds, the first 40 of which were excised to eliminate initialization effects for a total of 60 seconds of data. Each parameter set was repeated once with a new seed to ensure sufficient samples were taken to report data with high confidence. The only source of randomness in the simulation is the choice of PRB; thus a single, long simulation is sufficient to capture that randomness. ns-3 simulation runs produce traces of Physical Sidelink Control Channel (PSCCH) and Physical Sidelink Shared Channel (PSSCH) transmissions and receptions by all UEs (over slot and sub-channel used), that are post-processed to derive simulation estimates of $P_{COL}$ and associated sub-events. This is achieved by observing instances of UEs using identical PRBs and tracing the channel's history to determine the type of collision. $N_{C}$ is determined simultaneously as $P_{COL}$ by counting the number of PSCCH/PSSCH transmissions in the PRB during collisions. The PRR is derived by counting the number of successfully decoded PSSCH\footnote{In ns-3, a successful PSSCH decode requires the corresponding PSCCH to be decoded as well.} transmissions. In the case of $N_{Se} > 1$, more than one successful duplicate PSSCH decode counts as one for the PRR calculation. $N_{a}$ is derived from a separate trace from the SPS algorithm, which measures $N_{a}$ directly during each reselection event for each UE. In addition to the previously mentioned metrics, we tested the PRR as a function of distance up to the largest $N_{UE}$ tested (200 UEs result in 200 meters) to ensure the network was fully connected. Fig.~\ref{fig:PRRdependance} proves that PRR is uniform over all possible distances; thus, there is no dependence on distance. The ns-3 simulation was conducted on a PC running Ubuntu 20.04 with AMD Ryzen 9 3900X 12-Core Processor 3.79 GHz while the data processing and analytical models were implemented using Matlab R2023b.

\begin{figure*}[htbp]
\centering
\begin{minipage}[t]{0.45\textwidth}
\centering
\includegraphics[width=.98\textwidth]{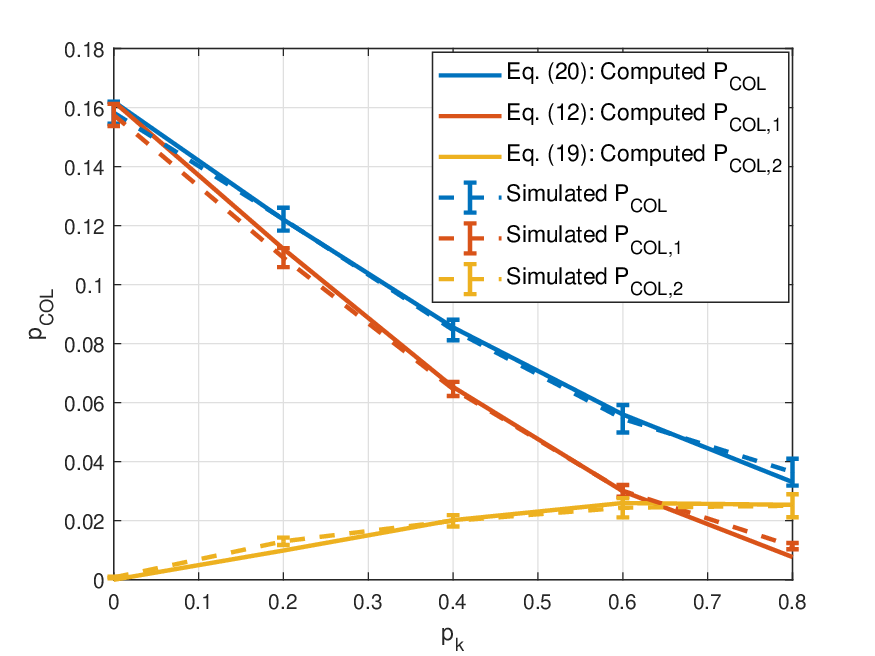}
\caption{$P_{COL} $ in terms of $p_k$: $N_{UE} = 100$.}
\label{fig:PcolvsPkNue100}
\end{minipage}
\begin{minipage}[t]{0.45\textwidth}
\centering
\includegraphics[width=.98\textwidth]{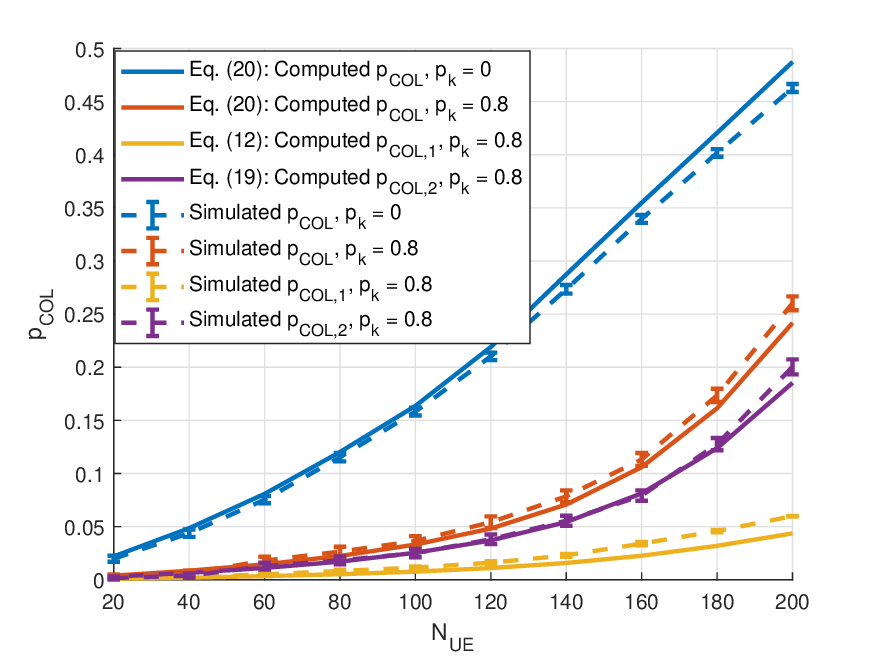}
\caption{$P_{COL} $ in terms of $N_{UE}$: $p_k = 0/0.8$.}
\label{fig:PcolvsPkvsNue}
\end{minipage}
\vspace{-0.5cm}
\end{figure*}

\subsection{Validation Results}
\underline{ $P_{COL} $ vs $p_k$, fixed $N_{UE}$ (Fig. \ref{fig:PcolvsPkNue100})}: 
 We observe that both $P_{COL} $ and $P_{COL,1} $ monotonically decreases with $p_k$ while $P_{COL,2} $ is monotonic increasing. 
The decrease of $P_{COL,1} $ can be explained by Eq. (\ref{eq:P_c1_complete}): when $p_k$ increases, UE 0 performs reselection less frequently in the selection window; meanwhile, the term $1- \left( 1 - \frac{2\pi_{0}(1-p_k)}{\overline{N_a }} \right)^{N_{UE}}$ also decreases, indicating that UE 0 is less likely to collide with other UEs given that UE 0 performs reselection in the selection window. Both effects cause $P_{COL,1} $ to decrease. The increase of $P_{COL,2} $ with $p_k$ can be explained as follows: when $p_k$ increases, UE 0 does not reselect more frequently in the selection window; meanwhile, if UE 0 collides with another UE in the selection window, their consecutive collisions last longer because they are more likely to stick with the previously reserved PRB. Since the decrease of $P_{COL,1} $ is much more significant than the increase of $P_{COL,2} $, $P_{COL} $ thereby increases with $p_k$. Note that the cross-over point between $P_{COL,1} $ and $P_{COL,2} $ occurs at $p_k \approx 0.64$, it follows that Collision Event 1 happens more frequently than Collision Event 2 in most $p_k$ cases (i.e., $p_k \in [0, 0.64]$). We conclude by remarking that the proposed models for $P_{COL} $, $P_{COL,1} $ and $P_{COL,2} $ matches well the ns-3 simulation results for a typical scenario ($N_{UE} = 100$).

\begin{figure}[t]
    \centering
\includegraphics[width=0.5\linewidth]{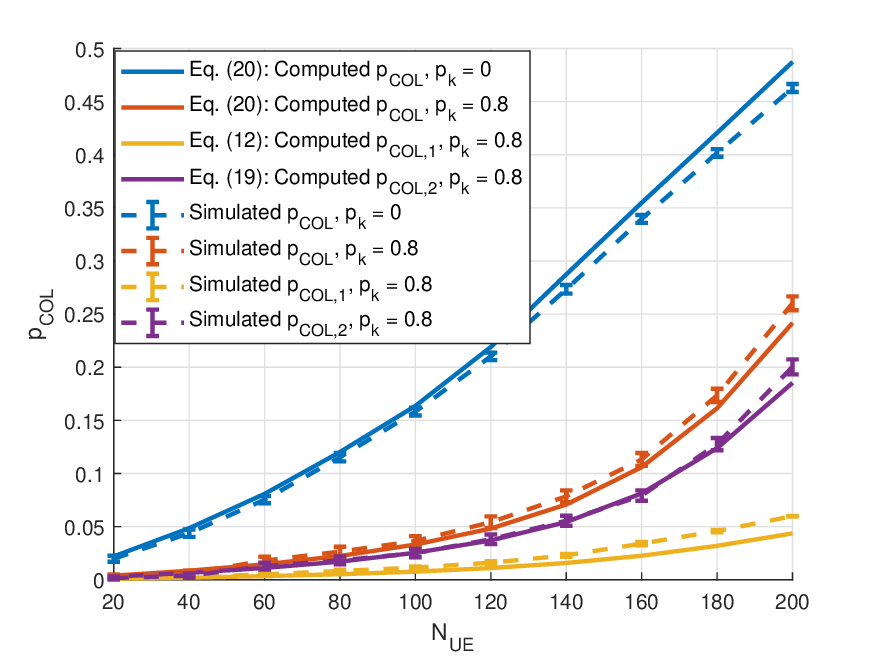}    \caption{$P_{COL} $ in terms of $N_{UE}$: $p_k = 0/0.8$.}
    \label{fig:PcolvsPkvsNue}
    \vspace{-0.5cm}
\end{figure}

\underline{$P_{COL} $ vs $N_{UE}$, varied $p_k$ (Fig. \ref{fig:PcolvsPkvsNue})}: Next, the computed $P_{COL} $, $P_{COL,1} $ and $P_{COL,2} $ overall matches the simulated ones at different $N_{UE}$. Note that when $p_k = 0$, only Collision Event 1 happens, thus $P_{COL} = P_{COL,1} $. When $N_{UE}$ goes up, both $P_{COL} $ under two $p_k$ go up. For $p_k = 0$, with increasing $N_{UE}$, more UEs participate in reselection, which leads $P_{COL} $ to increase. Hence for $p_k = 0.8$ and higher $N_{UE}$, likelihood of Collision Event 2 (which dominates $P_{COL} $) is higher, which leads $P_{COL} $ to increase. Compared with the simulation results, the analytical model subtly overestimates $P_{COL} $ with $p_k = 0$ at $N_{UE} \geq 160$. This gap can be explained by the $\overline{N_a }$ difference shown in Fig. \ref{fig:NavsPkvsNue}. The computed $\overline{N_a }$ is very close to the simulated values at low $N_{UE}$, however, the computed $\overline{N_a }$ deviates from  the simulated one around $N_{UE} \geq 160$. The lower $\overline{N_a }$ indicates that each UE has fewer available PRBs for reselection, and two UEs are more likely to re-select the same available PRB. Therefore, the computed $P_{COL} $ is overestimated compared to the simulated $P_{COL} $ at high $N_{UE}$. Note that the total number of PRBs in each RRI is $N_r = 200$ while the channel moves to saturation for $N_{UE} \geq 160$. Thus, the proposed analytical model can accurately predict $P_{COL} $ and $\overline{N_a }$ in an under-saturated condition while it becomes less accurate in a saturated condition ($ N_{UE}\approx N_r$). This is due to PHY issues that must be incorporated into the analysis for better accuracy under the saturated condition. For instance, large numbers of UEs in a dense local network imply that their respective RSRPs at a receiver are closer in value. As a result, when the RSRP threshold is increased to include more available PRBs for reselection, the actual $\overline{N_a }$ might be higher than the computed $\overline{N_a }$ in a saturated condition, rendering the pure MAC analysis less accurate. Note that saturated channel assumption is impractical for SL communications operationally, because the significant $PRR$ decrease will not satisfy the desired reliable requirements for critical messages.


\underline{$PRR $ vs $p_k$, fixed $N_{UE}$ (Fig. \ref{fig:PRRvsPkNue100})}: At $N_{UE} = 100$ (an under-saturated case), the computed $PRR $ matches quite well with the simulated one under different $p_k$, and $PRR $ increases as $p_k$ increases. \underline{$PRR $ vs $N_{UE}$,} \underline{varied $p_k$ (Fig. \ref{fig:PRRvsNuevsPk})}: Meanwhile, the computed $PRR $ also matches with the simulated one at different $N_{UE}$. Note that the computed $PRR $ with $p_k = 0$ at high $N_{UE}$ is subtly lower than the simulated one because the computed $P_{COL} $ with $p_k = 0$ is overestimated at high $N_{UE}$.
\begin{figure*}[htbp]
\centering
\begin{minipage}[t]{0.45\textwidth}
\centering
\includegraphics[width=.98\textwidth]{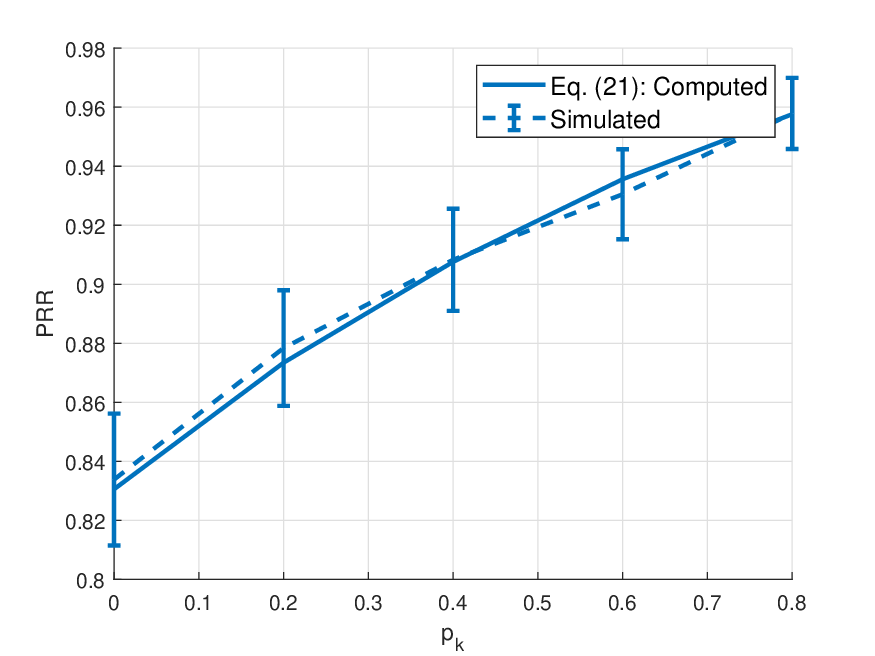}
\caption{$PRR $ in terms of $p_K$: $N_{UE} = 100$.}
\label{fig:PRRvsPkNue100}
\end{minipage}
\begin{minipage}[t]{0.45\textwidth}
\centering
\includegraphics[width=.98\textwidth]{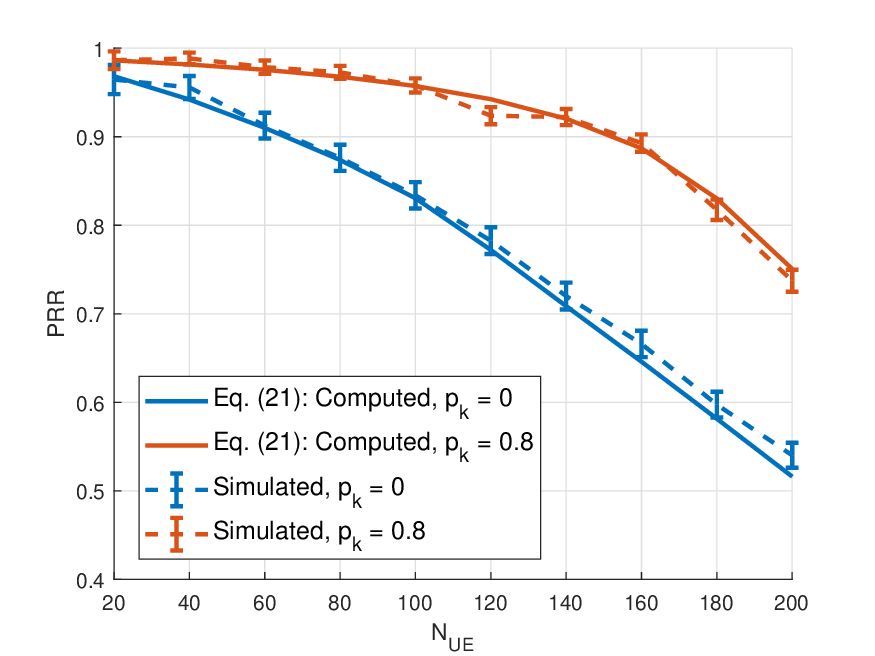}
\caption{$PRR $ in terms of $N_{UE}$: $p_k = 0/0.8$.}
\label{fig:PRRvsNuevsPk}
\end{minipage}
\end{figure*}


\underline{$P_{COL} (N_{Se})$ vs $N_{UE}$, fixed $p_k$ (Fig. \ref{fig:PcolvsNSevsNuePk0})}: The computed $P_{COL} (N_{Se})$ overall matches with the simulated $P_{COL} (N_{Se})$ at different $N_{UE}$, especially in the under-saturated condition. As expected, for a given $P_{COL} $ (e.g., $P_{COL} = 0.1$), $N_{UE} \, \approx \, 80$ for $N_{Se} = 1$ is almost twice of $N_{UE} \, \approx 40 $ for $N_{Se} = 2$. Meanwhile, at the lower $N_{UE}$ (e.g., $N_{UE} = 60$), $P_{COL} (2)$ ($\approx$ 0.2) is almost twice of $P_{COL} (1)$ ($\approx$ 0.1). Also, it is clear that $P_{COL} (2)$ is always higher than $P_{COL} (1)$ for all $N_{UE}$. 

\begin{figure*}[htbp]
\centering
\begin{minipage}[t]{0.45\textwidth}
\centering
\includegraphics[width=.98\textwidth]{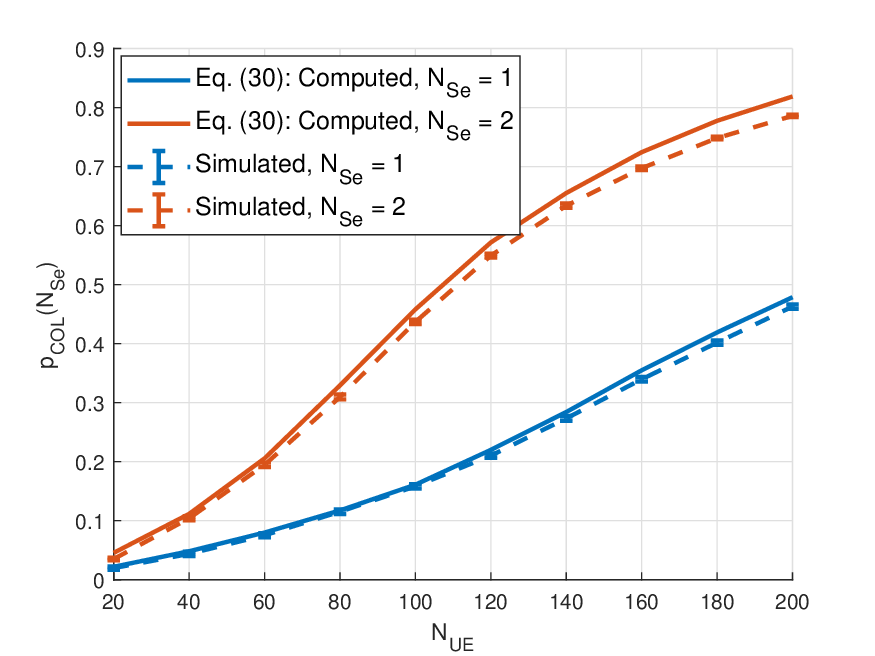}
\caption{$P_{COL} (N_{Se})$ in terms of $N_{UE}$: $p_k = 0$.}
\label{fig:PcolvsNSevsNuePk0}
\end{minipage}
\begin{minipage}[t]{0.45\textwidth}
\centering
\includegraphics[width=.98\textwidth]{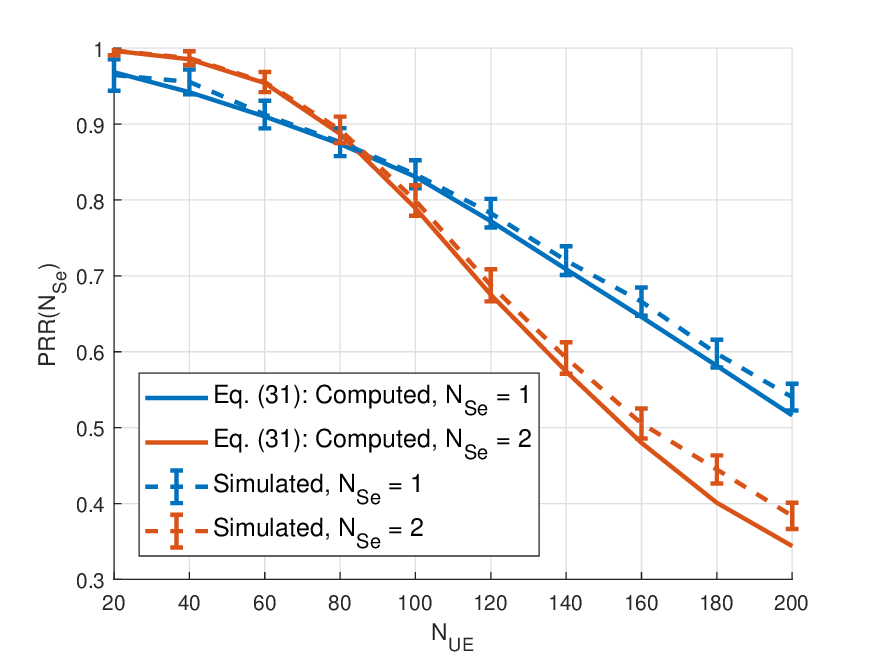}
\caption{$PRR (N_{Se})$ in terms of $N_{UE}$: $p_k = 0$.}
\label{fig:PRRvsNSevsvsNuePk0}
\end{minipage}
\end{figure*}


\underline{$PRR (N_{Se})$ vs $N_{UE}$, fixed $p_k$ (Fig. \ref{fig:PRRvsNSevsvsNuePk0})}: However, the $PRR (N_{Se})$ does not always follow the behavior of $P_{COL} (N_{Se})$ vs $N_{UE}$. For under-saturated condition, e.g., $N_{UE} \leq 80$, $PRR (2)$ is always higher than $PRR (1)$. Note that per Eq. (\ref{eq:Prr_multi}), $PRR (N_{Se})$ is proportional to $1 - \Bigl(P_{COL} (N_{Se})\Bigl)^{N_{Se}}$ if $P_{HD}$ is neglected ($N_{Se}$ does not affect $P_{HD}$), where $\Bigl(P_{COL} (N_{Se})\Bigl)^{N_{Se}}$ represents the probability that all $N_{Se}$ packets are not decoded. In the under-saturated condition, $N_{Se} = 2$ improves the packet reception gain because $P_{COL} (1) > \Bigl(P_{COL} (2)\Bigl)^{2}$. However, when $N_{UE} \geq 100$ (saturated scenario), $PRR (2)$ significantly dominates $PRR (1)$. Thus $PRR (N_{Se}=2)$ degrades significantly due to $P_{COL} (1) < \Bigl(P_{COL} (2)\Bigl)^{2}$. Note that the cross-over value ($N_{UE}^*, PRR (N_{Se}, N_{UE}^*)$) represents the performance boundary in terms of $N_{UE}$ and $N_{Se}$: $PRR (N_{Se}, N_{UE})$ benefits from the larger $N_{Se}$ if $N_{UE} < N_{UE}^*$, and degrades if $N_{UE} > N_{UE}^*$. $N_{UE}^*$ can be determined by solving $PRR (1, N_{UE}^*) = PRR (2, N_{UE}^*)$.

\begin{figure*}[tbp]
\centering
\begin{minipage}[t]{0.45\textwidth}
\centering
\includegraphics[width=.98\textwidth]{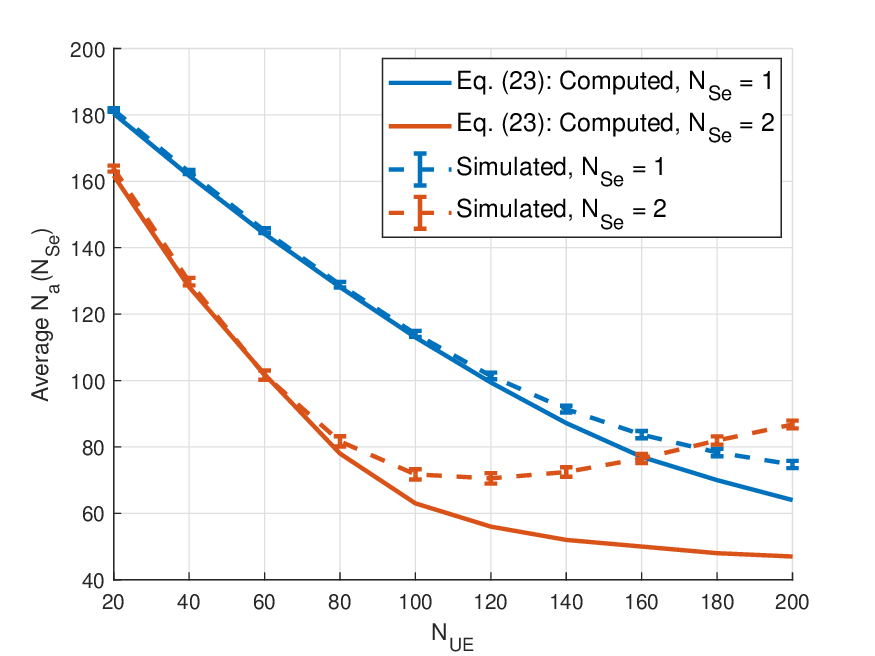}
\caption{$\overline{N_a }(N_{Se})$ in terms of $N_{UE}$: $p_k = 0$.}
\label{fig:NavsNSevsNuePk0}
\end{minipage}
\begin{minipage}[t]{0.45\textwidth}
\centering
\includegraphics[width=.98\textwidth]{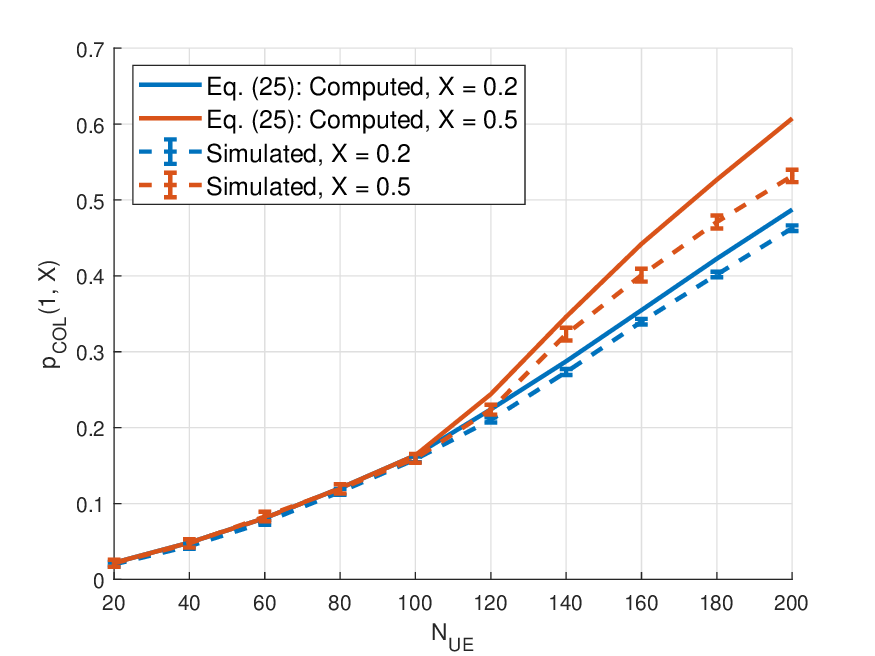}
\caption{$P_{COL} (1, X)$ in terms of $N_{UE}$: $p_k = 0$.}
\label{fig:PcolvsNuevsXPk0}
\end{minipage}
\vspace{-0.5cm}
\end{figure*}



\underline{$\overline{N_a }(N_{Se})$ vs $N_{UE}$, fixed $p_k$ (Fig. \ref{fig:NavsNSevsNuePk0})}: The computed $\overline{N_a }(1)$ is close to the simulated value when $N_{UE} < 160$ while $\overline{N_a }(2)$ matches with the simulated when $N_{UE} < 100$ representing the under-saturated condition ($N_{UE} \leq 160$ for $N_{Se} = 1$ and $N_{UE} \leq 100$ for $N_{Se} = 2$). Thus, computed $\overline{N_a }(N_{Se})$ is accurately in the under-saturated condition. In addition, the $\overline{N_a }(N_{Se})$ difference for $N_{Se} = 2$ becomes quite large in the saturated condition. For instance, when $N_{UE} = 200$ and $N_{Se} = 2$, the number of packets transmitted in each RRI is $N_{UE}N_{Se} = 400$, which is twice the total number of PRBs in each RRI $N_r = 200$. It should be noted that the simulated $\overline{N_a }(N_{Se})$ goes up as $N_{UE}$ increases in the saturated condition. This behavior occurs because the measured RSRP of each PRB becomes very close in the saturated condition, and some of the available PRBs for reselection become occupied. Increasing the RSRP threshold (e.g., 3dB granularity is increased each time) will include a burst of PRBs: the more the packets are transmitted in the saturated condition, the more the PRBs are included within each 3 dB granularity, and the higher the $\overline{N_a }(N_{Se})$ in the simulation is, which contradicts the fact in the model that more transmitted packets lead to the lower $\overline{N_a }(N_{Se})$. Thus, the pure MAC analysis becomes less accurate regarding $P_{COL} (N_{Se})$ and $PRR (N_{Se})$ in the saturated condition. 

\begin{figure*}[htbp]
\centering
\begin{minipage}[t]{0.45\textwidth}
\centering
\includegraphics[width=.98\textwidth]{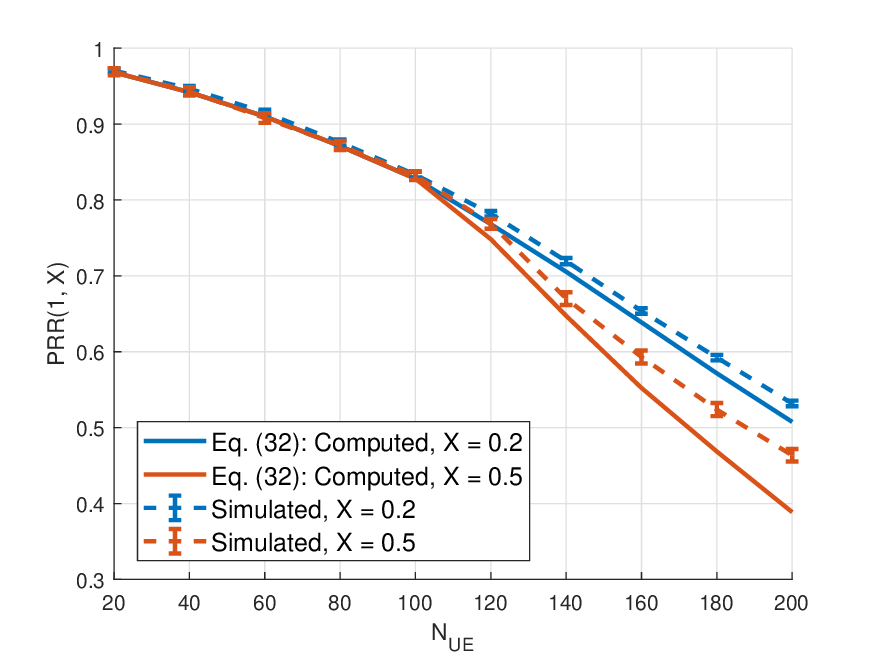}
\caption{$PRR(1,X)$ in terms of $N_{UE}$: $p_k = 0$.}
\label{fig:PRRvsNuevsXPk0}
\end{minipage}
\begin{minipage}[t]{0.45\textwidth}
\centering
\includegraphics[width=.98\textwidth]{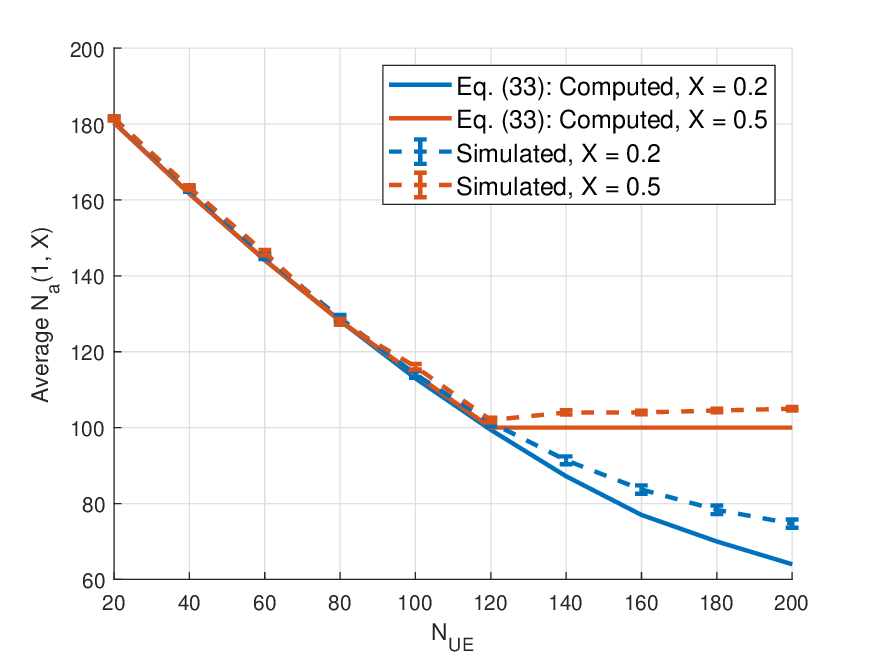}
\caption{$\overline{N_a}(1,X)$ in terms of $N_{UE}$: $p_k = 0$.}
\label{fig:NavsNuevsXPk0}
\end{minipage}
\end{figure*}


\underline{$P_{COL}(1, X)$ and $PRR(1, X)$ vs $N_{UE}$, fixed $p_k$ (Fig. \ref{fig:PcolvsNuevsXPk0} and \ref{fig:PRRvsNuevsXPk0})}: The computed $P_{COL} (1, X)$ and $PRR(1, X)$ are validated to predict well based on the simulated ones in the under-saturated channel condition. Meanwhile, $\overline{N_a}(1,X)$ in terms of $N_{UE}$ is validated in Fig. \ref{fig:NavsNuevsXPk0}. Note that, according to Eq. (\ref{eq:NaNseX}),  the minimum $\overline{N_a}(1,0.2) =0.2\cdot 200 = 40$ while the minimum $\overline{N_a}(1,0.5) = 0.5\cdot 200 = 100$. $\overline{N_a}(1)$ in both cases (only including unoccupied PRBs in available PRBs) must be the same if $\overline{N_a}(1) \geq 100$. Therefore, when  $N_{UE} \leq 100$, indicating that $\overline{N_a}(1) \geq 100$,  $P_{COL}(1, 0.5)$ (or $PRR(1, 0.5)$) should be the same as $P_{COL} (1, 0.2)$ (or $PRR(1, 0.2)$), which is validated in Fig. \ref{fig:PcolvsNuevsXPk0} (or Fig. \ref{fig:PRRvsNuevsXPk0}).  As $N_{UE}$ further increases, meaning that $\overline{N_a}(1) < 100$, $\overline{N_a}(1,0.5) = 100$ (including both unoccupied and occupied PRBs in available PRBs) remains fixed. By contrast, $\overline{N_a}(1,0.2)$ keeps decreasing because $\overline{N_a}(1) > 40$ still holds (only including unoccupied PRBs in available PRBs). As a result, $P_{COL}(1, 0.5)$ (or $PRR(1, 0.5)$) is no longer the same as $P_{COL} (1, 0.2)$ (or $PRR(1, 0.2)$) in such cases.


\section{Conclusion}
\label{conclusion}
In this work, we developed novel MAC collision models to thoroughly investigate the performance of the SPS protocol for 5G NR SL. Particularly, we defined the MAC collision events that accurately characterized the essence of MAC-based SPS. We also showed how the key SPS features relate to and further determine the MAC collision performance. The computed results were validated via simulations using the 5G-LENA module in ns-3. Through the extended analytical models, we illustrated that the MAC PRR benefits from the new feature - multiple Layer-2 packet transmission ($N_{Se}$) only in the under-saturated channel condition. Meanwhile, setting another feature - the minimum proportion of available PRBs for reselection ($X$) greater than 0.2 for the safety-related messages provides no benefit in any channel condition. Hence, we recommend that network engineers always set $X$ to the minimum allowable value and 3GPP consider lowering the current lower limit of $X = 0.2$ for SL reliability enhancement.

As the proposed pure MAC-based model does not work well in the saturated channel condition due to PHY issues, future work will adapt the current model by incorporating relevant PHY parameters to address the above problem. In addition, it would be interesting to validate if the proposed NR mode 2 MAC PRR model is also compatible with the LTE mode 4 under the same SPS configurations. 


\appendix
\subsection{$\text{Pr}\{R_{c,init}^{UE 0} < R_{c}^{UE 1}\}$ calculation}\label{cal1}
According to Fig. \ref{fig:MC_RC} and Eq. (\ref{eq:P_c22_c}), 
\begin{equation}
    \resizebox{.65\hsize}{!}{$
    \begin{aligned}
        &\text{Pr}\{R_{c,init}^{UE 0} < R_{c}^{UE 1}\} =  \sum_{|I|}\text{Pr}\{ R_{c, init}^{UE0} = i \}\cdot \text{Pr}\{R_{c, init}^{UE0}< R_{c}^{UE1}| R_{c, init}^{UE0} = i \}\\&=\text{Pr}\{\text{$R_c^{UE1} = 0$}|\text{$R_c^{UE0} = 0$ during UE 0 most recent selection window}\} \\& \cdot \text{Pr}\{\text{ $R_{c, init}^{UE1} > $}\text{$R_{c,init}^{UE0}$ during UE 0 most recent selection window}\} + \\& \text{Pr}\{\text{$R_c^{UE1} \neq 0$}|\text{$R_c^{UE0} = 0$ during UE 0 most recent selection window}\} \\& \cdot \text{Pr}\{\text{ $R_{c, init}^{UE1} > $}\text{$R_{c,init}^{UE0}$ during UE 0 most recent selection window}\}\\&= 2\pi_0\frac{5}{11} + (1-2\pi_0) \Big(\frac{1}{11}\cdot \frac{5}{11} + \frac{1}{11} \cdot \frac{45}{121}+ \frac{1}{11}\cdot\frac{36}{121}  \frac{1}{11} \cdot\frac{28}{121}+ \\&\frac{1}{11} \cdot \frac{21}{121} + \frac{1}{11} \cdot \frac{15}{121} + \frac{1}{11} \cdot \frac{10}{121} \frac{1}{11}\cdot\frac{6}{121}  + \frac{1}{11}\cdot \frac{3}{121}+ \frac{1}{11}\cdot \frac{1}{121}  + \\& \frac{1}{11}\cdot\frac{0}{121}\Big) = 0.2892\cdot2\pi_0 + 0.1653.\end{aligned}
        $}
    \end{equation}

\subsection{$\text{Pr}\{R_{c,init}^{UE 0} > R_{c}^{UE 1}\}$ calculation}\label{cal2}
According to Fig. \ref{fig:MC_RC} and Eq. (\ref{eq:P_c23_c}), 
\begin{equation}
    \resizebox{.65\hsize}{!}{$
    \begin{aligned}
        &\text{Pr}\{R_{c,init}^{UE 0} > R_{c}^{UE 1}\} =  \sum_{|I|}\text{Pr}\{ R_{c, init}^{UE0} = i \}\cdot \text{Pr}\{R_{c, init}^{UE0}> R_{c}^{UE1}| R_{c, init}^{UE0} = i \}\\&=\text{Pr}\{\text{$R_c^{UE1} = 0$}|\text{$R_c^{UE0} = 0$ during UE 0's most recent selection window}\} \\& \cdot \text{Pr}\{\text{ $R_{c, init}^{UE1} < $}\text{ $R_{c,init}^{UE0}$ during UE 0's most recent selection window}\} + \\& \text{Pr}\{\text{$R_c^{UE1} \neq 0$}|\text{$R_c^{UE0} = 0$ during UE 0's most recent selection window}\} \\& \cdot \text{Pr}\{\text{ $R_{c, init}^{UE1} < $}\text{ $R_{c,init}^{UE0}$ during UE 0's most recent selection window}\} \\&=2\pi_0\frac{5}{11} + (1-2\pi_0)\Big(\frac{1}{11}\cdot \frac{5}{11} + \frac{1}{11} \cdot \frac{6}{11}+ \frac{1}{11}\cdot\frac{76}{121}  + \frac{1}{11} \cdot\frac{85}{121}+ \\& \frac{1}{11} \cdot \frac{93}{121} + \frac{1}{11} \cdot \frac{100}{121}+ \frac{1}{11} \cdot \frac{106}{121} + \frac{1}{11}\cdot\frac{111}{121} + \frac{1}{11}\cdot \frac{115}{121}+ \frac{1}{11}\cdot \frac{118}{121}  + \\& \frac{1}{11}\cdot\frac{120}{121} \Big)= 0.7851 -0.3306\cdot2\pi_0.\end{aligned}
        $}
    \end{equation}

\bibliographystyle{IEEEtran}
\bibliography{reference}

@article{garcia2021tutorial,
  title={{A tutorial on 5G NR V2X communications}},
  author={Garcia, Mario H Casta{\~n}eda and Molina-Galan, Alejandro and Boban, Mate and Gozalvez, Javier and Coll-Perales, Baldomero and {\c{S}}ahin, Taylan and Kousaridas, Apostolos},
  journal={IEEE Communications Surveys \& Tutorials},
  volume={23},
  number={3},
  pages={1972--2026},
  year={2021},
  publisher={IEEE}
}

@inproceedings{wei2024optimized,
  title={{Optimized non-primary channel access design in IEEE 802.11 bn}},
  author={Wei, Dongyu and Cao, Liu and Zhang, Lyutianyang and Gao, Xiangyu and Yin, Hao},
  booktitle={Proc. the IEEE Global Communications Conference},
  pages={4588--4593},
  year={2024},
  month={Dec.},
  address={Cape Town, South Africa},
  organization={IEEE}
}

@techreport{3gpp.37.885,
 author = {3GPP},
 month = {Mar},
 title = {{S}tudy on {NR} {V}ehicle-to-{E}verything {(V2X)}},
 year = {2019},
 institution =  {The 3rd Generation Partnership Project {(3GPP)}},
 number = {{TR}37.885}
 }

@techreport{3gpp.37.985,
 author = {3GPP},
 month = {Jul},
 title = {Overall description of Radio Access Network {(RAN)}
aspects for Vehicle-to-everything {(V2X)} based on {LTE and NR}},
 year = {2020},
 institution =  {The 3rd Generation Partnership Project {(3GPP)}},
 number = {{TR}37.985}}

@techreport{3gpp.36.213,
 author = {3GPP},
 month = {Dec},
 title = {{Evolved Universal Terrestrial Radio Access (E-UTRA); Physical layer procedures}},
 year = {2020},
   institution =  {The 3rd Generation Partnership Project {(3GPP)}},
 number = {{TR}36.213},
}

@techreport{3gpp.36.321,
 author = {3GPP},
 month = {Dec},
 title = {Evolved Universal Terrestrial Radio Access {(E-UTRA)}; Medium Access Control {(MAC)} protocol specification},
 year = {2020},
   institution =  {The 3rd Generation Partnership Project {(3GPP)}},
 number = {{TR}36.321},
}

@article{weerackody2023needs,
  title={{Who Needs Basestations When We Have Sidelinks?}},
  author={Weerackody, Vijitha and Benson, Kent and Roy, Sumit},
  journal={Global Communications},
  volume={2023},
  year={2023}
}

@inproceedings{firstnet,
  author = {FirstNet},
  title = {{First Responder Network Authority}},
  year ={www.firstnet.gov}
}

@article{ali20213gpp,
  title={{3GPP NR V2X mode 2: overview, models and system-level evaluation}},
  author={Ali, Zoraze and Lag{\'e}n, Sandra and Giupponi, Lorenza and Rouil, Richard},
  journal={IEEE Access},
  volume={9},
  pages={89554--89579},
  year={2021},
  publisher={IEEE}
}

@inproceedings{brady2022modeling,
  title={{Modeling of NR C-V2X Mode 2 Throughput}},
  author={Brady, Collin and Cao, Liu and Roy, Sumit},
  booktitle={2022 IEEE International Workshop Technical Committee on Communications Quality and Reliability (CQR)},
  pages={19--24},
  year={2022},
  organization={IEEE}
}

@inproceedings{cecchini2017ltev2vsim,
  title={{LTEV2Vsim: An LTE-V2V simulator for the investigation of resource allocation for cooperative awareness}},
  author={Cecchini, Giammarco and Bazzi, Alessandro and Masini, Barbara M and Zanella, Alberto},
  booktitle={2017 5th IEEE International Conference on Models and Technologies for Intelligent Transportation Systems (MT-ITS)},
  pages={80--85},
  year={2017},
  organization={IEEE}
}

@inproceedings{mccarthy2019opencv2x,
  title={{OpenCV2X mode 4: A simulation extension for cellular vehicular communication networks}},
  author={McCarthy, Brian and O'Driscoll, Aisling},
  booktitle={2019 IEEE 24th International Workshop on Computer Aided Modeling and Design of Communication Links and Networks (CAMAD)},
  pages={1--6},
  year={2019},
  organization={IEEE}
}

@inproceedings{campolo20195g,
  title={{5G NR V2X: On the impact of a flexible numerology on the autonomous sidelink mode}},
  author={Campolo, Claudia and Molinaro, Antonella and Romeo, Francesco and Bazzi, Alessandro and Berthet, Antoine O},
  booktitle={2019 IEEE 2nd 5G World Forum (5GWF)},
  pages={102--107},
  year={2019},
  organization={IEEE}
}

@article{bazzi2018study,
  title={{Study of the impact of PHY and MAC parameters in 3GPP C-V2V mode 4}},
  author={Bazzi, Alessandro and Cecchini, Giammarco and Zanella, Alberto and Masini, Barbara M},
  journal={IEEE Access},
  volume={6},
  pages={71685--71698},
  year={2018},
  publisher={IEEE}
}

@inproceedings{xu2025enhanced,
  title={Enhanced SPS Velocity-Adaptive Scheme: Access Fairness in 5G NR V2I Networks},
  author={Xu, Xiao and Wu, Qiong and Fan, Pingyi and Wang, Kezhi},
  booktitle={2025 IEEE International Workshop on Radio Frequency and Antenna Technologies (iWRF\&AT)},
  pages={294--299},
  year={2025},
  organization={IEEE}
}

@ARTICLE{11054065,
  author={Bin Ali Wael, Chaeriah and Hadj Dogheche, El and Armi, Nasrullah and Subekti, Agus and Dayoub, Iyad},
  journal={IEEE Open Journal of Intelligent Transportation Systems}, 
  title={Leveraging 3GPP Features and Optimization Techniques for 5G NR-V2X Resource Allocation: A Survey}, 
  year={2025},
  volume={6},
  number={},
  pages={967-994},
  doi={10.1109/OJITS.2025.3584024}}

@inproceedings{nabil2018performance,
  title={{Performance analysis of sensing-based semi-persistent scheduling in C-V2X networks}},
  author={Nabil, Amr and Kaur, Komalbir and Dietrich, Carl and Marojevic, Vuk},
  booktitle={2018 IEEE 88th vehicular technology conference (VTC-Fall)},
  pages={1--5},
  year={2018},
  organization={IEEE}
}

@inproceedings{chen2021performance,
  title={{Performance evaluation of C-V2X mode 4 communications}},
  author={Chen, Miling and Chai, Rong and Hu, Hang and Jiang, Wenhang and He, Lin},
  booktitle={2021 IEEE Wireless Communications and Networking Conference (WCNC)},
  pages={1--6},
  year={2021},
  organization={IEEE}
}

@article{nba2020discrete,
  title={{A discrete-time Markov chain based comparison of the MAC layer performance of C-V2X mode 4 and IEEE 802.11 p}},
  author={NBA, Geeth P Wijesiri and Haapola, Jussi and Samarasinghe, Tharaka},
  journal={IEEE Transactions on Communications},
  volume={69},
  number={4},
  pages={2505--2517},
  year={2020},
  publisher={IEEE}
}

@INPROCEEDINGS{10757976,
  author={Lin, Kai-Yu and Wen, Chih-Yu},
  booktitle={2024 IEEE 100th Vehicular Technology Conference (VTC2024-Fall)}, 
  title={Packet Reception Analysis of C-V2X Mode 4 Communication for Highway Scenarios}, 
  year={2024},
  volume={},
  number={},
  pages={1-5},
  doi={10.1109/VTC2024-Fall63153.2024.10757976}}

@article{gonzalez2018analytical,
  title={{Analytical models of the performance of C-V2X mode 4 vehicular communications}},
  author={Gonzalez-Mart{\'\i}n, Manuel and Sepulcre, Miguel and Molina-Masegosa, Rafael and Gozalvez, Javier},
  journal={IEEE Transactions on Vehicular Technology},
  volume={68},
  number={2},
  pages={1155--1166},
  year={2018},
  publisher={IEEE}
}

@ARTICLE{10533255,
  author={Lusvarghi, Luca and Coll-Perales, Baldomero and Gozalvez, Javier and Merani, Maria Luisa},
  journal={IEEE Internet of Things Journal}, 
  title={Link Level Analysis of NR V2X Sidelink Communications}, 
  year={2024},
  volume={11},
  number={17},
  pages={28385-28397},
  doi={10.1109/JIOT.2024.3402551}}

@ARTICLE{11023842,
  author={Yáñez, Alexis and Salas, Felipe and Azurdia-Meza, Cesar A. and Ignacio Sandoval, Jorge and Céspedes, Sandra},
  journal={IEEE Access}, 
  title={Enhancing Urban Road Safety: A 5G NR Model for Vulnerable Road Users Awareness}, 
  year={2025},
  volume={13},
  number={},
  pages={99170-99182},
  doi={10.1109/ACCESS.2025.3576726}}

@article{gu2022performance,
  title={{Performance Analysis and Optimization for Semi-Persistent Scheduling in C-V2X}},
  author={Gu, Xin and Peng, Jun and Cai, Lin and Cheng, Yijun and Zhang, Xiaoyong and Liu, Weirong and Huang, Zhiwu},
  journal={IEEE Transactions on Vehicular Technology},
  volume={72},
  number={4},
  pages={4628--4642},
  year={2022},
  publisher={IEEE}
}

@inproceedings{gu2021performance,
  title={{Performance analysis on access collision in semi-persistent scheduling of C-V2X mode 4}},
  author={Gu, Xin and Peng, Jun and Cheng, Yijun and Zhang, Xiaoyong and Liu, Weirong and Huang, Zhiwu and Cai, Lin},
  booktitle={2021 IEEE 94th Vehicular Technology Conference (VTC2021-Fall)},
  pages={1--5},
  year={2021},
  organization={IEEE}
}

@article{liu2023towards,
  title={{Towards 5G new radio sidelink communications: A versatile link-level simulator and performance evaluation}},
  author={Liu, Peng and Shen, Chen and Liu, Chunmei and Cintr{\'o}n, Fernando J and Zhang, Lyutianyang and Cao, Liu and Rouil, Richard and Roy, Sumit},
  journal={Computer Communications},
  year={2023},
  publisher={Elsevier}
}

@inproceedings{rehman2022analytical,
  title={Analytical Modeling of Multiple Access Interference in {C-V2X} Sidelink Communications},
  author={Rehman, Abdul and Di Marco, Piergiuseppe and Valentini, Roberto and Santucci, Fortunato},
  booktitle={2022 IEEE International Mediterranean Conference on Communications and Networking (MeditCom)},
  pages={215--220},
  year={2022},
  organization={IEEE}
}

@article{rehman2023impact,
  title={On the Impact of Multiple Access Interference in {LTE-V2X} and {NR-V2X} Sidelink Communications},
  author={Rehman, Abdul and Valentini, Roberto and Cinque, Elena and Di Marco, Piergiuseppe and Santucci, Fortunato},
  journal={Sensors},
  volume={23},
  number={10},
  pages={4901},
  year={2023},
  publisher={MDPI}
}

\end{document}